\documentclass[12pt]{article}
\catcode`\@=11

\global\arraycolsep=2pt
\usepackage{amsbsy,amssymb,latexsym,amsfonts,amsmath}
\usepackage{graphicx,color,hyperref,braket}
\usepackage{tikz}
\usetikzlibrary{intersections, calc, arrows.meta}
\usepackage{graphicx}
\usepackage{braket}

\allowdisplaybreaks

\title{ADE triality via (non-)invertible symmetry gauging} 
\author{Yu Nakayama and Takahiro Tanaka}
\date{\today}

\begin{document}

\begin{flushright}
YITP-25-91

\end{flushright}

\vspace*{0.7cm}

\begin{center}
{\Large ADE triality via (non-)invertible symmetry gauging}
\vspace*{1.5cm}\\
{Takahilo Tanaka and Yu Nakayama}

\end{center}

\vspace*{1.0cm}
\begin{center}

Yukawa Institute for Theoretical Physics,
Kyoto University, Kitashirakawa Oiwakecho, Sakyo-ku, Kyoto 606-8502, Japan

\vspace{1.5cm}
\end{center}

\begin{abstract}
It is long known that A-series minimal models and D-series minimal models are exchanged by gauging the invertible $\mathbb{Z}_2$ symmetry. More recently, it has been shown that A-series minimal models and E-series minimal models are exchanged by gauging a non-invertible symmetry. We complete the triality picture by showing that D-series minimal models and E-series minimal models are exchanged by gauging another non-invertible symmetry. 
\end{abstract}

\thispagestyle{empty} 

\setcounter{page}{0}

\newpage

\tableofcontents

\newpage

\section{Introduction}

``What's done cannot be undone" \cite{Shao:2023gho}: the notion of non-invertible symmetry has been a paradigm shift in quantum field theories (QFTs) beyond the traditional notion of unitary symmetry advocated by Wigner in the 1930s. What's coming next? In this paper, we pursue a newer paradigm ``What's done can be undone" if we gauge the non-invertible symmetry: a network of non-invertible symmetries makes their gauging operation invertible.

Wigner claimed that the symmetry of a quantum system is specified by a group $G$. Then we can construct unitary operators $U_g(t)$ on a time slice labeled by elements $g$ of $G$, whose fusion rules are group-like (up to possible phases when projective). The paradigm shift in the past decade is to relax Wigner's assumption in various manners \cite{Gaiotto:2014kfa}. See \cite{Shao:2023gho, Davighi:2025iyk, McGreevy:2022oyu, Schafer-Nameki:2023jdn, Brennan:2023mmt, Iqbal:2024pee, Luo:2023ive, Bhardwaj:2023kri, Cordova:2022ruw} for reviews on generalized symmetry. In the modern perspective, symmetries are identified with topological defect $\mathcal{D}_x(M_p)$ with some labels $x$ on a $p$-dimensional submanifold $M_p$ of the spacetime. This new perspective gives a generalization of the conventional symmetries in QFT in two ways. The first is to allow more diverse co-dimension of the topological defects, leading to higher-form symmetries. The second is to consider fusion rules which are not group-like, but categorical. The latter type of symmetry is called non-invertible or categorical symmetry. 

In some more specific contexts, non-invertible symmetries have a longer history. In particular, non-invertible symmetries of two-dimensional conformal field theory (CFT) have been studied since the early days. The most famous example would be
the Kramers–Wannier duality defect of the critical Ising model  \cite{Bhardwaj:2017xup, Chang:2018iay, Verlinde:1988sn, Oshikawa:1996ww, Oshikawa:1996dj, Frohlich:2004ef, Frohlich:2006ch, Moore:1988qv, Thorngren:2019iar, Thorngren:2021yso, Huang:2021zvu, Burbano:2021loy, Lin:2022dhv, Chang:2022hud, Lu:2022ver, Lin:2023uvm, Nagoya:2023zky}. In more recent years, non-invertible symmetries have been discussed in higher dimensional QFTs such as four-dimensional massless Quantum Electrodynamics (QED) and axion-Maxwell theory \cite{Cordova:2023her, Putrov:2023jqi, Yamamoto:2023uzq, Buican:2023bzl, Cordova:2022fhg, Cordova:2022ieu, Choi:2022zal, Choi:2021kmx, Choi:2022fgx, Choi:2022jqy, Cordova:2024ypu, Yokokura:2022alv, Hidaka:2024kfx, Choi:2022rfe, Choi:2023pdp, Cordova:2023ent, DelZotto:2024ngj, Arbalestrier:2024oqg}, providing e.g. new constraints on the renormalization group flow and selection rules. 

The non-invertible symmetries can be useful even in well-established QFTs such as Virasoro minimal models. The Virasoro minimal model is a class of two-dimensional rational conformal field theory (RCFT), possessing many topological defect lines (TDLs). Virasoro minimal models are classified into three types in terms of ADE Dynkin diagrams, known as ADE classification \cite{Cappelli:1987xt, Cappelli:2009xj}. In a series of remarkable papers \cite{Petkova:2000ip, Petkova:2001ag}, Petkova and Zuber constructed all the TDLs of each Virasoro minimal model and investigated their fusion rules as well as twisted partition functions. 

What can we learn from the so-constructed TDLs?  Regarding TDLs as objects and TDL changing operators as morphisms, we can describe symmetries with the fusion category. As an example of the categorical structure, we can read the spin contents of the twisted Hilbert space $\mathcal{H}_{L_x}$ from the twisted partition function $Z_{L_x}$. The spin contents encode F-symbols of the symmetry category which are solutions of the pentagon identities. Because fusion categories cannot have non-trivial deformation, which is a fact known as Ocneanu rigidity, the spin contents of a TDL preserved under a relevant deformation is an RG invariant \cite{Chang:2018iay, Kikuchi:2021qxz, Kikuchi:2022rco}. Pursuing this idea, we have constrained RG flows between Virasoro minimal models \cite{Nakayama:2022svf} and proposed infinitely new ones \cite{Nakayama:2024msv}. These are consequences of ``what's done cannot be undone" beyond what can be obtained from Wigner's symmetry.

Let us now discuss ``what's done can be undone" when we gauge it. Suppose we are asked to gauge a finite group symmetry $G$. The procedure is to decorate a QFT $\mathcal{T}$ with TDLs $L_g(g \in G)$, which encode transition functions of a $G$-bundle, and sum over non-equivalent TDL configuration with appropriate weight. In two-dimensional CFTs, it is known as $G$-orbifolding \cite{DiFrancesco:1997nk}. Anomaly-free condition is translated into invariance under local deformation of TDL configuration. Now the paradigm shift here is that we can generalize the idea to non-invertible symmetry $\mathcal{C}$ cases \cite{Bhardwaj:2017xup,Frohlich:2009gb,Carqueville:2012dk,Brunner:2013xna, Diatlyk:2023fwf, Choi:2023vgk, Perez-Lona:2023djo,Perez-Lona:2024sds,Yu:2025iqf}. We first choose a TDL $\mathcal{A} \in \mathcal{C}$ and its multiplication morphism $m \in \mathrm{Hom}(\mathcal{A} \times \mathcal{A}, \mathcal{A})$ corresponding to the weight. By dressing a QFT with a mesh of $\mathcal{A}$, we can define the gauged theory $\mathcal{T}/\mathcal{A}$. 

A non-trivial property of gauging a group symmetry is that the gauged theory $\mathcal{T}/G$ has the dual symmetry $\hat{G}$ so that gauging $\hat{G}$ recovers the original theory $\mathcal{T}/G/\hat{G} \cong \mathcal{T}$. For example, we can exchange the A-series Virasoro minimal model and D-series Virasoro minimal models by gauging (dual) $\mathbb{Z}_2$ symmetry. Surprisingly this property also holds in non-invertible symmetry cases. The gauged theory $\mathcal{T}/\mathcal{A}$ has the dual line $\mathcal{A}^*$ such that $\mathcal{T}/\mathcal{A}/\mathcal{A^*} \cong \mathcal{T}$.
This is remarkable because each line cannot be inverted, but the network of lines can be invertible!
The goal of this paper is to demonstrate the idea ``what's done can be undone" to complete the ADE triality picture via non-invertible symmetry gauging.

The organization of our paper is as follows. In section \ref{TDL}, we give a review of the way to construct TDLs of Virasoro minimal models and see some examples with emphasis on D-series and E-series. 
In section \ref{noninvgauge}, we give the definition of gauging non-invertible symmetries and claim that $\mathcal{M}(A_{10}, D_{7})$ and $\mathcal{M}(A_{10}, E_{6})$ exchange each other via non-invertible symmetry gauging, hence completing the ADE triality. In section \ref{Conclusion}, we conclude with some discussions for future directions.

\section{Topological defect lines of Virasoro minimal models}\label{TDL}
\subsection{The topological defect lines in rational conformal field theory}
We consider a two-dimensional RCFT whose Hilbert space takes the form 
\begin{align}
    \mathcal{H} = \bigoplus_{j, \bar{j} \in \mathcal{I}} Z_{j\bar{j}}\mathcal{V}_j\otimes\bar{\mathcal{V}}_{\bar{j}},
\end{align}
where $\mathcal{I}$ is the finite set of the irreducible representations of the chiral algebra and $Z_{j\bar{j}}$ is the non-negative multiplicities. We denote $1$ as the trivial representation. A TDL $L$ in RCFT is topological in the sense 
\begin{align}
    [L_n,\hat{L}] = [\bar{L}_n, \hat{L}] = 0,\label{topologicalness}
\end{align}
where $\hat{L}$ is a topological defect line along the spacial direction and $L_n(\bar{L}_n)$ are (anti-)holomorphic Virasoro generators. The solutions of (\ref{topologicalness}) are linear combinations 
\begin{align}
    \hat{L}_x = \sum_{j, \bar{j}; \alpha, \alpha^{\prime}}\dfrac{\Psi_x^{(j, \bar{j}; \alpha, \alpha^{\prime})}}{\sqrt{S_{1j}S_{1\bar{j}}}} P^{(j, \bar{j}; \alpha, \alpha^{\prime})}\label{basischange}
\end{align}
of the projectors 
\begin{align}
    P^{(j, \bar{j}; \alpha, \alpha^{\prime})} : (\mathcal{V}_j\otimes\bar{\mathcal{V}}_{\bar{j}})^{(\alpha)} \to (\mathcal{V}_j\otimes\bar{\mathcal{V}}_{\bar{j}})^{(\alpha^{\prime})} \ \ \ \ \alpha, \alpha^{\prime} = 1, \cdots, Z_{j\bar{j}},
\end{align}
where $x$ is a label of TDL taking $n = \sum_{j, \bar{j}}(Z_{j\bar{j}})^2$ kinds of values and $S_{ij}$ is the modular S-matrix. The purpose of this change of the basis is to endow $\hat{L}_x$ with good properties such as simpleness and defect Hilbert space interpretation (i.e. Cardy-like condition).
For the identity line $L_1$, we set
\begin{align}
    \Psi_1^{(j, \bar{j}, \alpha, \alpha^{\prime})} = \sqrt{S_{1j}S_{1\bar{j}}}\delta_{\alpha\alpha^{\prime}}
\end{align}
and get
\begin{align}
    \hat{L}_1 = \sum_{j, \bar{j}, \alpha} P^{(j, \bar{j}, \alpha)} = 1.
\end{align}

When we insert two TDLs $L_x$ and $L_y$ along the time direction, they impose twisted boundary conditions and the spectrum changes into
\begin{align}
    \mathcal{H}_{y|x} = \bigoplus_{i, \bar{i} \in \mathcal{I}} {\widetilde{V}_{i\bar{i}^{*};x}}^{y}\mathcal{V}_{i}\otimes\bar{\mathcal{V}}_{\bar{i}},
\end{align}
where ${\widetilde{V}_{i\bar{i}^{*};x}}^{y}$ is the non-negative integer multiplicities. Then, the twisted partition function is 
\begin{align}
    Z_{y|x} = \mathrm{Tr}_{\mathcal{H}_{y|x}}q^{L_0 - \frac{c}{24}}\bar{q}^{\bar{L}_0 - \frac{c}{24}} = \sum_{i, \bar{i} \in \mathcal{I}} {\widetilde{V}_{i\bar{i}^{*};x}}^{y}\chi_{i}(\tau)\bar{\chi}_{\bar{i}}(\bar{\tau}),\label{sxypf}
\end{align}
where $\tau$ is the moduli of the spacetime torus, $q = e^{2\pi i\tau}$ and $\chi_i(\tau)$ is Virasoro character of the primary operator $\phi_i$. In order for the twisted partition function to have a Hilbert space interpretation, we require that $V_{i\bar{i}^{*};x}^{y}$ be a non-negative integer. Note that unlike in the untwisted partition function, the spin $L_0 - \bar{L}_0$ does not have to take (half) integer.

By performing modular S-transformation,\footnote{We use $\bar{\chi}_{\bar{j}}(\bar{\tilde{\tau} }) = \bar{S}_{\bar{j}\bar{i}} \bar{\chi}_{\bar{i}}(\bar{\tau}) $ and $\bar{S}_{\bar{j}\bar{i}} = S_{\bar{i}^* \bar{j}}$ in the charge conjugation invariant theories. The latter means $S_{1j}$ is real. In the Virasoro minimal models, we will eventually make no distinction between $i$ and $i^*$.} we can also express $Z_{y|x}$ as 
\begin{align}
    Z_{y|x} = \mathrm{Tr}_{\mathcal{H}} \left({\hat{L}_y}^{\dagger}\hat{L}_x\widetilde{q}^{L_0 - \frac{c}{24}}\bar{\widetilde{q}}^{\bar{L}_0 - \frac{c}{24}}\right) = \sum_{i, \bar{i}}\sum_{j, \bar{j}, \alpha, \alpha^{\prime}} \dfrac{S_{ij}S_{\bar{i}^*\bar{j}}}{S_{1j}S_{1\bar{j}}}\Psi_{x}^{(j, \bar{j}, \alpha, \alpha^{\prime})}\overline{\Psi_y^{(j, \bar{j}, \alpha, \alpha^{\prime})}}\chi_{i}(\tau)\bar{\chi}_{\bar{i}}(\bar{\tau}),\label{txypf}
\end{align}
where $\widetilde{\tau} = -\tau^{-1}$. Comparing (\ref{sxypf}) and (\ref{txypf}) leads to
\begin{align}
    {\widetilde{V}_{i\bar{i};x}}^{y} = \sum_{j, \bar{j}, \alpha, \alpha^{\prime}} \dfrac{S_{ij}S_{\bar{i}\bar{j}}}{S_{1j}S_{1\bar{j}}}\Psi_{x}^{(j, \bar{j}, \alpha, \alpha^{\prime})}\overline{\Psi_y^{(j, \bar{j}, \alpha, \alpha^{\prime})}}.\label{Vtilde}
\end{align}
Here, following the ansatz made in \cite{Petkova:2000ip, Petkova:2001ag,Petkova:2001zn, Petkova:2013yoa}, we assume that (\ref{basischange}) is a unitary change of basis from $\{P^{(j, \bar{j}, \alpha, \alpha^{\prime})}\}$ to $\{\hat{L}_x\}$.\footnote{The authors do not know if relaxing the assumption may or may not give us more consistent TDLs. In the examples below, we believe we have found all the consistent TDLs whose maximal number is bounded by $n$.}  Then using (\ref{Vtilde}) and Verlinde formula
\begin{align}
    N_{ij}^{k} = \sum_{l} \dfrac{S_{il}S_{jl}\bar{S}_{lk}}{S_{1l}},\label{vf}
\end{align}
$n\times n$ matrices $\widetilde{V}_{ij}$ whose entries are $(\widetilde{V}_{ij})_{xy} = {\widetilde{V}_{i\bar{i};x}}^{y}$ satisfy
\begin{align}
    \widetilde{V}_{i_1j_1}\widetilde{V}_{i_2j_2} = \sum_{i_3, j_3} N_{i_1i_2}^{i_3}N_{j_1j_2}^{j_3}\widetilde{V}_{i_3j_3}.\label{VVNNV}
\end{align}
Thus we can rephrase that finding TDLs corresponds to solving the non-linear non-negative integer matrix equation (\ref{VVNNV}) (under the ansatz that the basis change $\Psi$ is unitary). This is called nimrep $\widetilde{V}$ of Double Fusion Algebra.
\subsection{Graph fusion algebra}
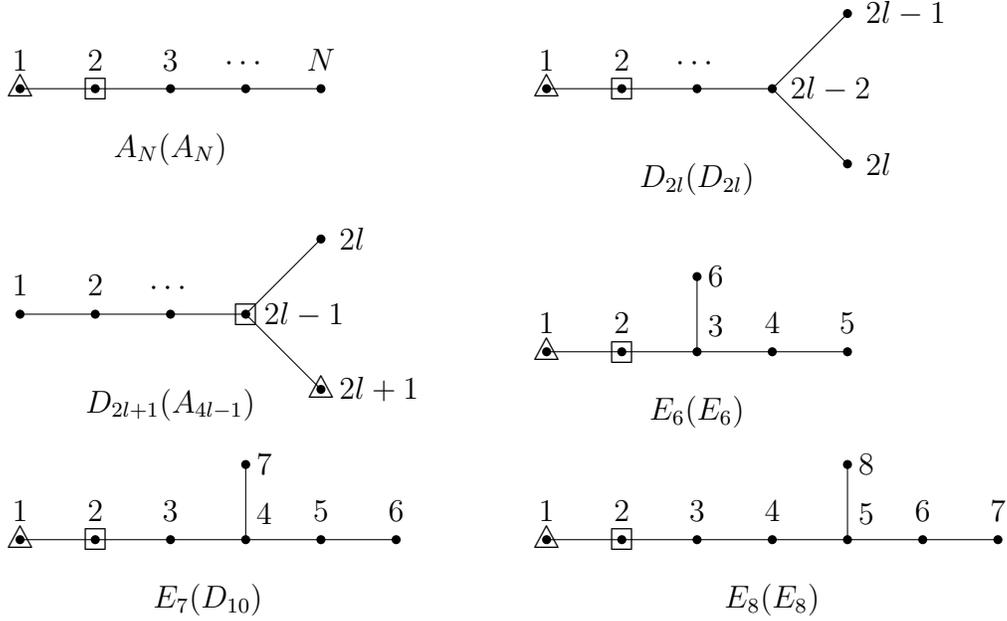
\begin{figure}[ht]
\centering
\begin{tikzpicture}
\draw(0,0)--(4,0);
\fill(0,0)circle(0.06);
\fill(1,0)circle(0.06);
\fill(2,0)circle(0.06);
\fill(3,0)circle(0.06);
\fill(4,0)circle(0.06);
\draw(0,0.1)node[above]{$1$};
\draw(1,0.1)node[above]{$2$};
\draw(2,0.1)node[above]{$3$};
\draw(3,0.1)node[above]{$\cdots$};
\draw(4,0.1)node[above]{$N$};
\draw(0,0)node{$\triangle$};
\draw(1,0)node{$\square$};
\draw(2,-0.8)node{$A_N(A_N)$};  
\draw(7,0)--(10,0);
\draw(11,1)--(10,0)--(11,-1);
\fill(7,0)circle(0.06);
\fill(8,0)circle(0.06);
\fill(9,0)circle(0.06);
\fill(10,0)circle(0.06);
\fill(11,1)circle(0.06);
\fill(11,-1)circle(0.06);
\draw(7,0)node{$\triangle$};
\draw(8,0)node{$\square$};
\draw(7,0.1)node[above]{$1$};
\draw(8,0.1)node[above]{$2$};
\draw(9,0.1)node[above]{$\cdots$};
\draw(10.1,0)node[right]{$2l - 2$};
\draw(11.1,1)node[right]{$2l - 1$};
\draw(11.1,-1)node[right]{$2l$};
\draw(9,-1.2)node{$D_{2l}(D_{2l})$};
\draw(0,-3)--(3,-3);
\draw(4,-2)--(3,-3)--(4,-4);
\fill(0,-3)circle(0.06);
\fill(1,-3)circle(0.06);
\fill(2,-3)circle(0.06);
\fill(3,-3)circle(0.06);
\fill(4,-2)circle(0.06);
\fill(4,-4)circle(0.06);
\draw(4,-4)node{$\triangle$};
\draw(3,-3)node{$\square$};
\draw(0,-2.9)node[above]{$1$};
\draw(1,-2.9)node[above]{$2$};
\draw(2,-2.9)node[above]{$\cdots$};
\draw(3.1,-3)node[right]{$2l - 1$};
\draw(4.1,-2)node[right]{$2l$};
\draw(4.1,-4)node[right]{$2l + 1$};
\draw(2,-4.2)node{$D_{2l + 1}(A_{4l - 1})$};
\draw(7,-3.5)--(11,-3.5);
\draw(9, -2.5)--(9, -3.5);
\fill(7,-3.5)circle(0.06);
\fill(8,-3.5)circle(0.06);
\fill(9,-3.5)circle(0.06);
\fill(10,-3.5)circle(0.06);
\fill(11,-3.5)circle(0.06);
\fill(9,-2.5)circle(0.06);
\draw(7,-3.4)node[above]{$1$};
\draw(8,-3.4)node[above]{$2$};
\draw(9,-3.15)node[right]{$3$};
\draw(10,-3.4)node[above]{$4$};
\draw(11,-3.4)node[above]{$5$};
\draw(9,-2.5)node[right]{$6$};
\draw(9,-4.3)node{$E_6(E_6)$};
\draw(7,-3.5)node{$\triangle$};
\draw(8,-3.5)node{$\square$};
\draw(0,-6)--(5, -6);
\draw(3, -5)--(3, -6);
\fill(0,-6)circle(0.06);
\fill(1,-6)circle(0.06);
\fill(2,-6)circle(0.06);
\fill(3,-6)circle(0.06);
\fill(4,-6)circle(0.06);
\fill(5,-6)circle(0.06);
\fill(3,-5)circle(0.06);
\draw(0,-5.9)node[above]{$1$};
\draw(1,-5.9)node[above]{$2$};
\draw(2,-5.9)node[above]{$3$};
\draw(3,-5.65)node[right]{$4$};
\draw(4,-5.9)node[above]{$5$};
\draw(5,-5.9)node[above]{$6$};
\draw(3,-5)node[right]{$7$};
\draw(2.5,-6.8)node{$E_7(D_{10})$};
\draw(0,-6)node{$\triangle$};
\draw(1,-6)node{$\square$};
\draw(7,-6)--(13, -6);
\draw(11, -5)--(11, -6);
\fill(7,-6)circle(0.06);
\fill(8,-6)circle(0.06);
\fill(9,-6)circle(0.06);
\fill(10,-6)circle(0.06);
\fill(11,-6)circle(0.06);
\fill(12,-6)circle(0.06);
\fill(13,-6)circle(0.06);
\fill(11,-5)circle(0.06);
\draw(7,-5.9)node[above]{$1$};
\draw(8,-5.9)node[above]{$2$};
\draw(9,-5.9)node[above]{$3$};
\draw(10,-5.9)node[above]{$4$};
\draw(11,-5.65)node[right]{$5$};
\draw(12,-5.9)node[above]{$6$};
\draw(13,-5.9)node[above]{$7$};
\draw(11,-5)node[right]{$8$};
\draw(10,-6.8)node{$E_8(E_8)$};
\draw(7,-6)node{$\triangle$};
\draw(8,-6)node{$\square$};
\end{tikzpicture}
\caption{ADE graphs. We also denote the parent of each graph in parentheses. The unit and fundamental node are shown by $\triangle$ and $\square$ respectively.}
\label{ADE graphs}
\end{figure}

Before we give the solutions of (\ref{VVNNV}), let us introduce the graph fusion algebra \cite{Behrend:1999bn, Chui:2002bp}. Consider an ADE graph $G$, the Dynkin diagram of simply laced Lie algebra, as in Figure \ref{ADE graphs}. 

We label the nodes of the graph $G$ whose Coxeter number is $g$ as $a,b, \cdots$ and use the same letter $G = (G_{ab})$ for the adjacency matrix. We denote its orthonormal eigenvectors for the eigenvalues 
\begin{align}
    \lambda_m = 2\cos\pi\dfrac{m}{g}
\end{align}
as $\psi_{a}^{m}$ and fix their phases by demanding $\psi_{\triangle}^m >0$ for all $m$. The set of $m$'s is called the exponents $\mathrm{Exp}(G)$. In Table \ref{GgExp}, we list $g$ and $\mathrm{Exp}(G)$ for each graph.
\begin{table}[ht]
\begin{center}
\begin{tabular}{c|c|c}
    $G$ & $g$ & $\mathrm{Exp}(G)$ \\ \hline
    $A_N$ & $N + 1$ & $1,2, \cdots, N$\\ \hline
    $D_N$ & $2N - 2$ & $1, 3, \cdots, 2N - 3, N - 1$\\ \hline
    $E_6$ & $12$ & $1, 4, 5, 7, 8, 11$\\ \hline
    $E_7$ & $18$ & $1, 5, 7, 9, 11, 13, 17$\\ \hline
    $E_8$ & $30$ & $1, 7, 11, 13, 17, 19, 23, 29$
\end{tabular}
\caption{Values of $g$ and $\mathrm{Exp}(G)$ for ADE graphs.}
\label{GgExp}
\end{center}
\end{table}

The graph fusion algebra is defined by
\begin{align}
    a \times b = \sum_{c}\hat{N}_{ab}^{(G)c}c,
\end{align}
where the integer structure constants are given by the Verlinde-like formula\footnote{We should note that it is not obvious at all the above definition of $\psi_a^m$ gives the integers $\hat{N}_{ab}^{(G)c}$ here but it turns out to be the case.}
\begin{align}
    \hat{N}_{ab}^{(G)c} = \sum_{m \in \mathrm{Exp}(G)} \dfrac{\psi_a^{m}\psi_{b}^{m}\overline{\psi_{c}^{m}}}{\psi_{\triangle}^{m}}.\label{Nhat}
\end{align}
In the case of $G = A_N$, $\psi_{a}^{m} = S_{am} = \sqrt{\frac{2}{g}}\sin\frac{am\pi}{g}$ and (\ref{Nhat}) reduces to the Verlinde formula (\ref{vf}).

For a graph $G$ whose Coxeter number is $g$, there is another algebra of the fused adjacency matrices $n_i^{(G)}$ whose structure constant is $N_{ij}^{(A_{g - 1})k}$. The fused adjacency matrices $n_i^{(G)}$ are defined recursively
\begin{align}
    n_1^{(G)} = I,\ \ n_2^{(G)} = G,\ \ n_{i + 1}^{(G)} = n_2^{(G)}n_i^{(G)} - n_{i - 1}^{(G)}\ (i = 3, \cdots, g - 2).
\end{align}
They can be also expressed by the Verlinde-like formula 
\begin{align}
    (n_i^{(G)})_a^{b} = \sum_{m \in \mathrm{Exp}(G)} \dfrac{S_{i}^{j}}{S_{\triangle}^{j}}\psi_{a}^{m}\overline{\psi_{b}^{m}},
\end{align}
where this S-matrix is that of $SU(2)_{g - 2}$ i.e. $S_{ij} = \sqrt{\frac{2}{g}}\sin\frac{ij\pi}{g}$. These matrices satisfy
\begin{align}
    n_i^{(G)}n_j^{(G)} = \sum_{k} N_{ij}^{(A_{g - 1})k}n_k^{(G)}.
\end{align}
In the following subsections, we will give the formula for $\widetilde{V}_{ij; x}^{y}$ in terms of $\hat{N}^{(G)}$ and $n_i^{(G)}$.
\subsection{The topological defect lines of Virasoro minimal models and the twisted partition functions}\label{TDLs}
Since $\widetilde{V}_{ij; x}^{y}$ represents the spectrum under the insertion of two TDLs $L_{x}$ and $L_y$ along the time direction, to determine $\widetilde{V}_{ij; x}^{y}$ is equivalent to determine $\widetilde{V}_{ij; z}^{1}$ and the fusion rules of TDLs 
\begin{align}
    L_{x} \times L_{y} = \sum_{z} \widetilde{N}_{xy}^{z}L_{z},
\end{align}
where we denote the fusion coefficients as $\widetilde{N}_{xy}^{z}$ so that $ \widetilde{V}_{ij; x}^{y} = \widetilde{N}_{xy}^{z} \widetilde{V}_{ij; z}^{1}$.

For general Virasoro minimal models $\mathcal{M}(A_{h - 1}, G)$, the seminal work \cite{Petkova:2000ip, Petkova:2001ag, Chui:2002bp} showed that the labels of TDLs take the form $x = (r, X) = (r, a, b, \kappa) \in (A_{h - 1}, H, H, \mathbb{Z}_2)$, where $H$ is the parent of $G$. See Figure \ref{ADE graphs} for the parent $H$ of each graph $G$.  

The fusion rules of TDLs are governed by the Ocneanu graph fusion algebra of $G$ whose structure constants are $N_{XY}^{(\mathrm{Oc}G)Z}$. The fusion rule of the TDLs in the Virasoro minimal model assumes the independent structures on $A_{h-1}$ labels and $G$ labels:
\begin{align}
    L_{(r, X)} \times L_{(r^{\prime}, X^{\prime})} = \sum_{r^{\prime\prime}, X^{\prime\prime}} N_{rr^{\prime}}^{(A_{h - 1})r^{\prime\prime}}N_{XX^{\prime}}^{(\mathrm{Oc}G)X^{\prime\prime}}L_{(r^{\prime\prime}, X^{\prime\prime})},\ \widetilde{N}_{xy}^{z} = N_{rr^{\prime}}^{(A_{h - 1})r^{\prime\prime}}N_{XX^{\prime}}^{(\mathrm{Oc}G)X^{\prime\prime}}.
\end{align}
In the end, $N_{XY}^{(\mathrm{Oc}G)Z}$ can be determined from the graph fusion matrix $\hat{N}_{ab}^{(G)c}$ introduced in the previous subsection, but the precise relation between $\hat{N}_{ab}^{(G)c}$ and $N_{XY}^{(\mathrm{Oc}G)Z}$ depends on the graph $G$, so we will see some examples separately for various $G$'s. 

Let us turn our attention to $\widetilde{V}_{ij; z}^{1}$. The starting point is the ADE classification of the modular invariant partition functions of the Virasoro minimal models \cite{Cappelli:1987xt, Cappelli:2009xj}. The modular invariant partition functions are
\begin{align}
    Z_{\mathcal{M}(A_{h - 1}, G)}(\tau) = \sum_{(r, s), (r^{\prime}, s^{\prime})} Z_{(r, s)(r^{\prime}, s^{\prime})}\chi_{(r, s)}(\tau)\bar{\chi}_{(r^{\prime}, s^{\prime})}(\bar{\tau})\ \ \ \ Z_{(r, s)(r^{\prime}, s^{\prime})} = \delta_{rr^{\prime}}\sum_{a \in T}n_{s1}^{(H)a}n_{s^{\prime}1}^{(H)\zeta(a)},
\end{align}
where $n_i^{(H)}$ are the fused adjacency matrices of the parent $H$,
\begin{align}
    T = T_1 = \begin{cases}
              \{1, 2, \cdots, g - 1\}\ \ \ \ \ \ G = A_{g - 1}\\
              \{1, 3, \cdots, 2l - 1, 2l\}\ \ \ \ G = D_{2l}\\
              \{1, 2, \cdots, 4l - 1\}\ \ \ \ \ \ G = D_{2l + 1}\\
              \{1, 5, 6\}\ \ \ \ G = E_6\\
              \{1, 3, 5, 7, 9, 10\}\ \ \ \ G = E_7\\
              \{1, 7\}\ \ \ \ G = E_8,
    \end{cases}
\end{align}
and $\zeta$ is the involution  
\begin{align}
    \zeta : \begin{cases}
        s \mapsto s\ \ \ \mathrm{for\ odd}\ s\\
        s \mapsto 4l - s\ \ \ \mathrm{for\ even}\ s
    \end{cases}\label{twist}
\end{align}
for $G = D_{2l + 1}$ case,
\begin{align}
    \zeta : \{1, 3, 5, 7, 9, 10\} \mapsto\ \{1, 9, 5, 7, 3, 10\}
\end{align}
for $G = E_7$ case and the identity map for the other cases.

We want to know the twisted partition function with the insertion of a TDL $L_{(r, a, b, \kappa)}$ along the time direction. 
In \cite{Petkova:2000ip, Petkova:2001ag, Chui:2002bp, OttoChui:2005qi, Coquereaux:2001di}, 
they showed that the following twisted partition functions satisfy all the constraints and hence determine $\widetilde{V}_{ij; z}^{1}$: 
\begin{align}
    Z_{(r, a, b, \kappa)}(\tau) = \sum_{(r^{\prime}, s^{\prime}), (r^{\prime\prime}, s^{\prime\prime})} N_{rr^{\prime}}^{(A_{h - 1})r^{\prime\prime}}[P_{ab}^{\kappa}]_{s^{\prime}s^{\prime\prime}}\chi_{(r^{\prime}, s^{\prime})}(\tau)\bar{\chi}_{(r^{\prime\prime}, s^{\prime\prime})}(\bar{\tau})\ \ \ \ [P_{ab}^{\kappa}]_{s^{\prime}s^{\prime\prime}} = \sum_{c \in T_{\kappa}} n_{s^{\prime}a}^{(H)c}n_{s^{\prime\prime}b}^{(H)\zeta(c)},\label{twistedpartition}
\end{align}
where 
\begin{align}
    T_2 = \begin{cases}
        \{2, 4, \cdots, 2l - 2\}\ \ \ \ G = D_{2l}\\
        T_1\ \ \ \ \mathrm{oterwise}
    \end{cases}
\end{align}
except for $G = E_7$ case. However, different labels do not necessarily give different TDLs. Two TDLs are identified when they give the same twisted partition functions
\begin{align}
    Z_{(r, a, b, \kappa)} = Z_{(r^{\prime}, a^{\prime}, b^{\prime}, \kappa^{\prime})} \Rightarrow L_{(r, a, b, \kappa)} = L_{(r^{\prime}, a^{\prime}, b^{\prime}, \kappa^{\prime})}.
\end{align}
This is because the TDLs giving the same twisted partition function act on states in the same way by the modular S-transformation.
In particular, it suffices for $x$ to take the form
\begin{align}
    x = \begin{cases}
        (r, s) \in A_{h - 1} \times A_{g - 1}\ \ \ \ \ \ G = A_{g - 1}\\
        (r, a, \kappa) \in A_{h - 1} \times D_{2l} \times \mathbb{Z}_2\ \ \ \ G = D_{2l}\\
        (r, s) \in A_{h - 1} \times A_{4l - 1}\ \ \ \ \ \ G = D_{2l + 1}\\
        (r, a, b) \in A_{h - 1} \times E_6 \times \{1, 2\}\ \ \ \ G = E_6\\
        (r, a, b) \in A_{h - 1} \times E_8 \times \{1, 2, 3, 8\}\ \ \ \ G = E_8,
        \end{cases}
\end{align}
together with the identification under the Kac symmetry
\begin{equation}
    \begin{split}
        L_{(r, s)} &= L_{(h - r, g - s)}\ \ \ \ G = A_{g - 1}\\
        L_{(r, X)} &= L_{(h - r, X)}\ \ \ \ \ \ \mathrm{otherwise}.\label{Lidentify}
    \end{split}
\end{equation}

Now we have obtained $\widetilde{V}_{ij; z}^{1}$, and the remaining task is to determine the fusion coefficients $\widetilde{N}_{xy}^z$. As we have mentioned, it is determined from the Ocneanu graph fusion algebra, but we have to resort to the case-by-case analysis \cite{Petkova:2000ip, Petkova:2001ag, Coquereaux:2001di}. Here we show some relevant examples necessary for our study.
\subsubsection{Example 1 : \texorpdfstring{$\mathcal{M}(A_{h - 1}, A_{g - 1})$}{TEXT}}\label{exAA}
The Ocneanu graph algebra of $A_{g - 1}$ is the same as the graph fusion algebra of $A_{g - 1}$, so its elements are labeled by the node $s$ of the graph $A_{g - 1}$ and $N_{ss}^{(\mathrm{Oc}A_{g - 1})s^{\prime\prime}} = N_{ss^{\prime}}^{(A_{g - 1})s^{\prime\prime}}$. Under the identification (\ref{Lidentify}), the diagonal minimal model $\mathcal{M}(A_{h - 1}, A_{g - 1})$ has $\frac{(h - 1)(g - 1)}{2}$ TDLs $L_{(r, s)} = L_{(h - r, g - s)}\ (1 \le r \le h - 1,\ 1 \le s \le g - 1)$ and their fusion rules are
\begin{align}
    L_{(r, s)} \times L_{(r^{\prime}, s^{\prime})} = \sum_{r^{\prime\prime},\ s^{\prime\prime}} N_{rr^{\prime}}^{(A_{h - 1})r^{\prime\prime}}N_{ss^{\prime}}^{(A_{g - 1})s^{\prime\prime}}L_{(r^{\prime\prime}, s^{\prime\prime})},
\end{align} 
which are the same as those of the primary operators. They are identified with the Verlinde lines \cite{Verlinde:1988sn} whose number is the same as the number of primary operators. They are consistent with the statement that there is one-to-one correspondence between TDLs and the primary operators in any diagonal RCFT.
\subsubsection{Example 2 : \texorpdfstring{$\mathcal{M}(A_{4}, D_{4})$(3-state Potts model)}{TEXT}}
The Ocneanu graph fusion algebra of $D_4$ is spanned by 8 elements labeled by $X = (a, \kappa) \in D_{4} \times \mathbb{Z}_2$ and 
\begin{align}
    N_{X}^{(\mathrm{Oc}D_{4})} = \begin{cases}
        \begin{pmatrix}
            \hat{N}_{a}^{(D_4)}&0\\0&\hat{N}_{a^*}^{(D_4)}
        \end{pmatrix}
        \ \ \ \ \mathrm{if}\ \kappa = 1\\
        \begin{pmatrix}
            0&\hat{N}_a^{(D_4)}\\ \hat{N}_{a^*}^{(D_4)}&0
        \end{pmatrix}
        \ \ \ \ \mathrm{if}\ \kappa = 2 ,
    \end{cases}
\end{align}
where $*$ operation exchanges $3$ and $4$ and leaves $1$ and $2$ invariant. The basis is given by 
\begin{equation}
    \begin{split}
        &L_{(1, 1, 1)} = 1,\ L_{(1, 2, 1)} = \mathcal{N}, \ L_{(1, 3, 1)} = \eta,\ L_{(1, 4, 1)} = \eta^2,\\ 
        &L_{(1, 1, 2)} = C,\ L_{(1, 2, 2)} = \mathcal{N}C,\ L_{(1, 3, 2)} = \eta C,\ L_{(1, 4, 2)} = \eta^2 C.
    \end{split}
\end{equation}

Under the identification (\ref{Lidentify}), there are two TDLs $1$ and $L_{(2, 1, 1)} = W$ from $A_4$ part. Combining them, $\mathcal{M}(A_{4}, D_{4})$, 3-state Potts model, has 16 TDLs and their fusion rules are \cite{Chang:2018iay, Gu:2023yhm}
\begin{align}
    &\eta^3 = 1,\ \mathcal{N} \times \eta = \eta \times \mathcal{N} = \mathcal{N},\ \mathcal{N}^2 = 1 + \eta + \eta^2,\notag\\
    &C^2 = 1,\ \eta \times C = C \times \eta^2,\ \eta^2 \times C = C \times \eta,\label{fusion3state}\\
    & W^2 = 1 + W.\notag
\end{align}
As you can see in the second line of (\ref{fusion3state}), the fusion of TDLs is non-commutative, which is the characteristic of $G = D_{2l}$. This is because $N_X^{(\mathrm{Oc}D_4)}$ is not a symmetric matrix. The first line of (\ref{fusion3state}) is the fusions of the $\mathbb{Z}_3$ Tambara-Yamagami category and the third line is those of the Lee-Yang category. $1, \eta$ and $C$ generate the invertible $S_3$ symmetry.

This model contains $12$ primary operators and has multiplicity $2$ at $(h, \bar{h}) = (\frac{1}{15}, \frac{1}{15}), (\frac{2}{3}, \frac{2}{3})$. Therefore the label of the TDL can take
\begin{align}
    n = \sum_{j, \bar{j}}(Z_{j\bar{j}})^2 = 8 \times 1^2 + 2 \times 2^2 = 16
\end{align}
kinds of values. It is consistent with the statement that $\mathcal{M}(A_4, D_4)$ has $16$ TDLs. Among the examples we look at in this paper, this is the only model with non-trivial multiplicities, so we have confirmed that $n$ and the number of TDLs match. Since the multiplicities of all the other examples are 1, we will not explicitly check this, but we can easily see that $n$ and the number of TDLs always match. 
\subsubsection{Example 3 : \texorpdfstring{$\mathcal{M}(A_{10}, D_{7})$}{TEXT}}\label{D7TDL}
The Ocneanu graph fusion algebra of $D_7$ is spanned by $11$ elements. Under the identification (\ref{Lidentify}), $\mathcal{M}(A_{10}, D_7)$ has 55 TDLs $L_{(r, s)} = L_{(11 - r, s)}\ (1 \le r \le 10,\ 1 \le s \le 11)$ and their fusion rules are the same as those of its parent theory $\mathcal{M}(A_{10}, A_{11})$
\begin{align}
    L_{(r, s)} \times L_{(r^{\prime}, s^{\prime})} = \sum_{r^{\prime\prime},\ s^{\prime\prime}} N_{rr^{\prime}}^{(A_{10})r^{\prime\prime}}N_{ss^{\prime}}^{(A_{11})s^{\prime\prime}}L_{(r^{\prime\prime},s^{\prime\prime})}.
\end{align}
The difference between $\mathcal{M}(A_{h - 1}, A_{g - 1})$ and $\mathcal{M}(A_{h - 1}, D_{g/2 + 1})$, where $g \in 4\mathbb{Z}$, is the twisted partition functions
\begin{align}
    Z_{(r, s)} = \begin{cases}
        \displaystyle\sum_{(r^{\prime}, s^{\prime}), (r^{\prime\prime}, s^{\prime\prime})} N_{rr^{\prime}}^{(A_{h - 1})r^{\prime\prime}}N_{ss^{\prime}}^{(A_{g - 1})s^{\prime\prime}}\chi_{(r^{\prime}, s^{\prime})}(\tau)\bar{\chi}_{(r^{\prime\prime}, s^{\prime\prime})}(\bar{\tau})\ \ \ \ \mathcal{M}(A_{h - 1}, A_{g - 1})\\
        \displaystyle\sum_{(r^{\prime}, s^{\prime}), (r^{\prime\prime}, s^{\prime\prime})} N_{rr^{\prime}}^{(A_{h - 1})r^{\prime\prime}}N_{ss^{\prime}}^{(A_{g - 1})\zeta(s^{\prime\prime})}\chi_{(r^{\prime}, s^{\prime})}(\tau)\bar{\chi}_{(r^{\prime\prime}, s^{\prime\prime})}(\bar{\tau})\ \ \ \ \mathcal{M}(A_{h - 1}, D_{g/2 + 1}).
    \end{cases}
\end{align}
\subsubsection{Example 4 : \texorpdfstring{$\mathcal{M}(A_{10}, E_{6})$}{TEXT}}\label{E6TDL}
The Ocneanu graph fusion algebra of $E_6$ is spanned by $12$ elements labeled by $X = (a, b) \in E_6 \times \{1, 2\}$ and 
\begin{align}
    N_{X}^{(\mathrm{Oc}E_6)} = \begin{cases}
        \begin{pmatrix}
            \hat{N}_{a}^{(E_6)}&0\\0&\hat{N}_{a}^{(E_6)}
        \end{pmatrix}
        \ \ \ \ \mathrm{if}\ x = (a, 1)\\
        \begin{pmatrix}
            0&\hat{N}_a^{(E_6)}\\ \hat{N}_{a}^{(E_6)}&\hat{N}_{a}^{(E_6)}\hat{N}_{6}^{(E_6)}
        \end{pmatrix}
        \ \ \ \ \mathrm{if}\ x = (a, 2).
    \end{cases}
\end{align}
See also \cite{Pearce:2024udz} for a concise expression.
The basis is given by 
\begin{equation}
    \begin{split}
           L_{(1, 1, 1)} = 1,\ &L_{(1, 2, 1)} = \mathcal{D}_1,\ L_{(1, 4, 1)} = \mathcal{D}_2,\ L_{(1, 3, 1)} = \mathcal{D}_3,\ L_{(1, 5, 1)} = \eta,\ L_{(1, 6, 1)} = \mathcal{N},\\ 
           &L_{(1, 1, 2)} = \mathcal{D}_{1}^{\prime}, \ L_{(1, 5, 2)} = \mathcal{D}_{2}^{\prime},\ L_{(1, 6, 2)} = \mathcal{D}_{3}^{\prime},\\ 
           &L_{(1, 2, 2)} = \mathcal{D}_1\mathcal{D}_{1}^{\prime} = \mathcal{D}_2\mathcal{D}_{2}^{\prime},\ L_{(1, 4, 2)} = \mathcal{D}_1\mathcal{D}_{2}^{\prime} = \mathcal{D}_{1}^{\prime}\mathcal{D}_{2}\\
           &L_{(1, 3, 2)} = \mathcal{D}_1\mathcal{D}_{3}^{\prime} = \mathcal{D}_2\mathcal{D}_{3}^{\prime} = \mathcal{D}_3\mathcal{D}_{1}^{\prime} = \mathcal{D}_3\mathcal{D}_{2}^{\prime}. \label{E6TDLs}
    \end{split}
\end{equation}

Under the identification (\ref{Lidentify}), there are five TDLs from $A_{10}$ part. Combining them, $\mathcal{M}(A_{10}, E_6)$ has 60 TDLs and their fusion rules among the TDLs without $^\prime$ are 
\begin{align}
    \mathcal{D}_1^2 = \mathcal{D}_2^2 = 1 + \mathcal{D}_3,\ \mathcal{D}_3^2 =  1 + \eta + 2\mathcal{D}_3,\ \mathcal{D}_1\eta = \mathcal{D}_2,\ \mathcal{D}_1\mathcal{N} = \mathcal{D}_3,\ \mathcal{N}^2 = 1 + \eta\label{E6fusion}. 
\end{align}
The fusion among the TDLs with $^\prime$ has the same structure as (\ref{E6fusion}). $\{1, \eta, \mathcal{N}\}$ forms $\mathbb{Z}_2$ Tambara-Yamagami category. $\{1,\ \eta,\ \mathcal{D}_3\}$ forms the $\frac{1}{2}E_6$ category and $\{1, \mathcal{D}_1, \mathcal{D}_2, \mathcal{D}_3, \eta, \mathcal{N}\}$ is a $\mathbb{Z}_2$-extension $\mathcal{C}_6$ of $\frac{1}{2}E_6$ category. $\{1, \mathcal{D}_1^{\prime}, \mathcal{D}_2^{\prime}, \mathcal{D}_3^{\prime}, \eta, \mathcal{N}\}$ is related to $\mathcal{C}_6$ through the parity flip \cite{Chang:2018iay, Diatlyk:2023fwf}. 
\section{Non-invertible symmetry gauging in \texorpdfstring{$\mathcal{M}(A_{10}, D_7)$}{TEXT} and \texorpdfstring{$\mathcal{M}(A_{10}, E_6)$}{TEXT}}\label{noninvgauge}
\subsection{Algebra and its constraint}
Before giving the definition of gauging non-invertible symmetries, let us recall how to gauge a non-anomalous finite group symmetry $G$ of a theory $\mathcal{T}$ on the torus. In the CFT literature, it is also known as the orbifolding of the $G$ symmetry.
By definition, coupling $\mathcal{T}$ with $G$-valued background gauge fields and summing over the inequivalent configurations of the gauge fields with appropriate weights give the gauged theory $\mathcal{T}/G$. In terms of TDLs, background gauging corresponds to inserting a network of $L_g(g \in G)$ along the dual triangulation of the torus which encodes the transition function of a $G$-bundle. Summing over all inequivalent configurations of the mesh gives the gauged theory. Moreover, there is a freedom to put a complex number $\varphi(g, h)$ and $\varphi^{\vee}(g, h) = \frac{1}{\varphi(g, h)}$ on the vertices among $L_g, L_h$ and $L_{gh}$, where $[\varphi] \in H^{2}(G, U(1))$
\begin{align}
    \begin{tikzpicture}[scale=1.0,baseline={([yshift=-.5ex]current bounding box.center)},vertex/.style={anchor=base,
    circle,fill=black!25,minimum size=18pt,inner sep=2pt},scale=0.50]        
    \draw[->](-2,-2)--(-1,-1);
    \draw[->](2,-2)--(1,-1);
    \draw[->](0,0)--(0,1);
    \draw(-1,-1)--(0,0);
    \draw(0,1)--(0,2);
    \draw(1,-1)--(0,0);
    \draw(0,0.5)node[right]{$\varphi(g, h)$};
    \draw(-2,-2)node[below]{$L_g$};
    \draw(2,-2)node[below]{$L_h$};
    \draw(0,2)node[above]{$L_{gh}$};
    \fill(0,0)circle(0.1);
    \draw(4,-2)node{$,$};
    \draw[->](8,-2)--(8,-1);
    \draw(8,-1)--(8,0);
    \draw[->](8,0)--(7,1);
    \draw(7,1)--(6,2);
    \draw[->](8,0)--(9,1);
    \draw(9,1)--(10,2);
    \draw(8,-0.5)node[right]{$\varphi^{\vee}(g, h) = \frac{1}{\varphi(g, h)}$};
    \draw(8,-2)node[below]{$L_{gh}$};
    \draw(6,2)node[above]{$L_g$};
    \draw(10,2)node[above]{$L_h$};
    \fill(8,0)circle(0.1);    
    \draw(16,-2)node{$.$};
    \end{tikzpicture}
\end{align}
This is called the discrete torsion. 

At the partition function level, we can write the gauged theory diagrammatically
\begin{equation}
    Z_{\mathcal{T}/G} = \dfrac{1}{\lvert G\rvert}\sum_{\substack{g, h \in G\\ gh = hg}} \varphi(g, h)\varphi^{\vee}(h, g)\ \ \ \begin{tikzpicture}[scale=0.6,baseline={([yshift=-.5ex]current bounding box.center)},vertex/.style={anchor=base,
    circle,fill=black!25,minimum size=18pt,inner sep=2pt},scale=0.50]        
    \draw(0,0)--(6, 0)--(6,6)--(0,6)--(0,0);
    \draw[red,->](3,0)--(3.5,1);
    \draw[red](3.5,1)--(4,2);
    \draw[blue,->](6,3)--(5,2.5);
    \draw[blue](5,2.5)--(4,2);
    \draw[green,->](4,2)--(3,3);
    \draw[green](3,3)--(2,4);
    \draw[blue,->](2,4)--(1,3.5);
    \draw[blue](1,3.5)--(0,3);
    \draw[red,->](2,4)--(2.5,5);
    \draw[red](2.5,5)--(3,6);
    \draw[red](2.5,5)node[right]{$L_g$};
    \draw[blue](5,2.5)node[above]{$L_h$};
    \draw[green](2.3,2.3)node{$L_{gh}$};
    \end{tikzpicture}
\end{equation}
where we introduce the normalization factor $\lvert G\rvert$. However, there is an alternative way to view this procedure which can be generalized to the cases of non-invertible symmetries. Consider the non-simple TDL $\mathcal{A} = \sum_{g \in G} L_g$ and insert a fine-enough trivalent mesh of $\mathcal{A}$. Moreover, we set the multiplication (co-)morphism $m$ and $m^{\vee}$ on the vertices among three $\mathcal{A}$'s to
\begin{align}
    \begin{tikzpicture}[scale=1.0,baseline={([yshift=-.5ex]current bounding box.center)},vertex/.style={anchor=base,
    circle,fill=black!25,minimum size=18pt,inner sep=2pt},scale=0.50]        
    \draw[->](-2,-2)--(-1,-1);
    \draw[->](2,-2)--(1,-1);
    \draw[->](0,0)--(0,1);
    \draw(-1,-1)--(0,0);
    \draw(0,1)--(0,2);
    \draw(1,-1)--(0,0);
    \draw(0,0.5)node[right]{$m$};
    \draw(-2,-2)node[below]{$\mathcal{A}$};
    \draw(2,-2)node[below]{$\mathcal{A}$};
    \draw(0,2)node[above]{$\mathcal{A}$};
    \fill(0,0)circle(0.1);
    \end{tikzpicture}
    &= \dfrac{1}{\sqrt{\lvert G\rvert}}\sum_{g, h \in G} 
    \begin{tikzpicture}[scale=1.0,baseline={([yshift=-.5ex]current bounding box.center)},vertex/.style={anchor=base,
    circle,fill=black!25,minimum size=18pt,inner sep=2pt},scale=0.50]
    \draw[->](-2,-2)--(-1,-1);
    \draw[->](2,-2)--(1,-1);
    \draw[->](0,0)--(0,1);
    \draw(-1,-1)--(0,0);
    \draw(0,1)--(0,2);
    \draw(1,-1)--(0,0);
    \draw(0,0.5)node[right]{$m_{g\ h}^{gh} = \varphi(g, h)$};
    \draw(-2,-2)node[below]{$L_g$};
    \draw(2,-2)node[below]{$L_h$};
    \draw(0,2)node[above]{$L_{gh}$};
    \fill(0,0)circle(0.1);      
    \end{tikzpicture}
    ,\notag\\
    \begin{tikzpicture}[scale=1.0,baseline={([yshift=-.5ex]current bounding box.center)},vertex/.style={anchor=base,
    circle,fill=black!25,minimum size=18pt,inner sep=2pt},scale=0.50]        
    \draw[->](8,-2)--(8,-1);
    \draw(8,-1)--(8,0);
    \draw[->](8,0)--(7,1);
    \draw(7,1)--(6,2);
    \draw[->](8,0)--(9,1);
    \draw(9,1)--(10,2);
    \draw(8,-0.5)node[right]{$m^{\vee}$};
    \draw(8,-2)node[below]{$\mathcal{A}$};
    \draw(6,2)node[above]{$\mathcal{A}$};
    \draw(10,2)node[above]{$\mathcal{A}$};
    \fill(8,0)circle(0.1);    
    \end{tikzpicture}
    &= \dfrac{1}{\sqrt{\lvert G\rvert}}\sum_{g, h \in G} 
    \begin{tikzpicture}[scale=1.0,baseline={([yshift=-.5ex]current bounding box.center)},vertex/.style={anchor=base,
    circle,fill=black!25,minimum size=18pt,inner sep=2pt},scale=0.50]
    \draw[->](8,-2)--(8,-1);
    \draw(8,-1)--(8,0);
    \draw[->](8,0)--(7,1);
    \draw(7,1)--(6,2);
    \draw[->](8,0)--(9,1);
    \draw(9,1)--(10,2);
    \draw(8,-0.5)node[right]{$(m^{\vee})_{g\ h}^{gh} = \varphi^{\vee}(g, h)$};
    \draw(8,-2)node[below]{$L_{gh}$};
    \draw(6,2)node[above]{$L_g$};
    \draw(10,2)node[above]{$L_h$};
    \fill(8,0)circle(0.1);    
    \draw(16,-2)node{$.$};
    \end{tikzpicture}
\end{align}
Because $L_g, L_h$ and $L_k$ cannot form a trivalent vertex if $k \neq gh$, we can see that the two viewpoints are equivalent. 

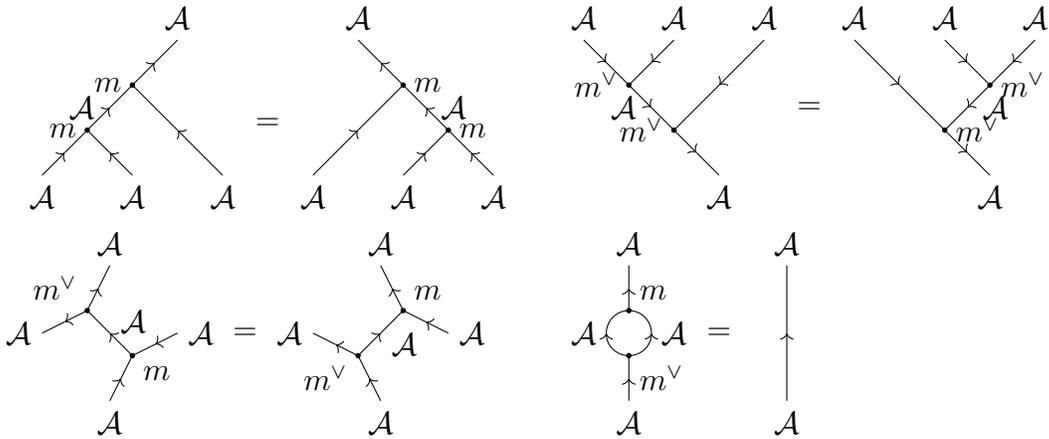
\begin{figure}[ht]
\centering
\begin{tikzpicture}[scale = 0.6]
\draw[->](0, 0)--(0.5, 0.5);
\draw(0.5, 0.5)--(1, 1);
\draw[->](2, 0)--(1.5, 0.5);
\draw(1.5, 0.5)--(1, 1);
\draw[->](1, 1)--(1.5, 1.5);
\draw(1.5, 1.5)--(2, 2);
\draw[->](4, 0)--(3, 1);
\draw(3, 1)--(2, 2);
\draw[->](2, 2)--(2.5, 2.5);
\draw(2.5, 2.5)--(3, 3);
\fill(1, 1)circle(0.06);
\fill(2, 2)circle(0.06);
\draw(1, 1)node[left]{$m$};
\draw(2, 2)node[left]{$m$};
\draw(0, 0)node[below]{$\mathcal{A}$};
\draw(2, 0)node[below]{$\mathcal{A}$};
\draw(4, 0)node[below]{$\mathcal{A}$};
\draw(3, 3)node[above]{$\mathcal{A}$};
\draw(1.4, 1.5)node[left]{$\mathcal{A}$};
\draw(5, 1.5)node[below]{$=$};
\draw[->](6, 0)--(7, 1);
\draw(7, 1)--(8, 2);
\draw[->](8, 0)--(8.5, 0.5);
\draw(8.5, 0.5)--(9, 1);
\draw[->](10, 0)--(9.5, 0.5);
\draw(9.5, 0.5)--(9, 1);
\draw[->](9, 1)--(8.5, 1.5);
\draw(8.5, 1.5)--(8, 2);
\draw[->](8, 2)--(7.5, 2.5);
\draw(7.5, 2.5)--(7, 3);
\fill(9, 1)circle(0.06);
\fill(8, 2)circle(0.06);
\draw(9, 1)node[right]{$m$};
\draw(8, 2)node[right]{$m$};
\draw(6, 0)node[below]{$\mathcal{A}$};
\draw(8, 0)node[below]{$\mathcal{A}$};
\draw(10, 0)node[below]{$\mathcal{A}$};
\draw(7, 3)node[above]{$\mathcal{A}$};
\draw(8.6, 1.5)node[right]{$\mathcal{A}$};
\draw[->](12, 3)--(12.5, 2.5);
\draw(12.5, 2.5)--(13, 2);
\draw[->](14, 3)--(13.5, 2.5);
\draw(13.5, 2.5)--(13, 2);
\draw[->](13, 2)--(13.5, 1.5);
\draw(13.5, 1.5)--(14, 1);
\draw[->](16, 3)--(15, 2);
\draw(15, 2)--(14, 1);
\draw[->](14, 1)--(14.5, 0.5);
\draw(14.5, 0.5)--(15, 0);
\fill(14, 1)circle(0.06);
\fill(13, 2)circle(0.06);
\draw(14, 1)node[left]{$m^{\vee}$};
\draw(13, 2)node[left]{$m^{\vee}$};
\draw(12, 3)node[above]{$\mathcal{A}$};
\draw(14, 3)node[above]{$\mathcal{A}$};
\draw(16, 3)node[above]{$\mathcal{A}$};
\draw(15, 0)node[below]{$\mathcal{A}$};
\draw(13.4, 1.5)node[left]{$\mathcal{A}$};
\draw(17, 1.5)node{$=$};
\draw[->](18, 3)--(19, 2);
\draw(19, 2)--(20, 1);
\draw[->](20, 3)--(20.5, 2.5);
\draw(20.5, 2.5)--(21, 2);
\draw[->](22, 3)--(21.5, 2.5);
\draw(21.5, 2.5)--(21, 2);
\draw[->](21, 2)--(20.5, 1.5);
\draw(20.5, 1.5)--(20, 1);
\draw[->](20, 1)--(20.5, 0.5);
\draw(20.5, 0.5)--(21, 0);
\fill(20, 1)circle(0.06);
\fill(21, 2)circle(0.06);
\draw(20, 1)node[right]{$m^{\vee}$};
\draw(21, 2)node[right]{$m^{\vee}$};
\draw(18, 3)node[above]{$\mathcal{A}$};
\draw(20, 3)node[above]{$\mathcal{A}$};
\draw(22, 3)node[above]{$\mathcal{A}$};
\draw(21, 0)node[below]{$\mathcal{A}$};
\draw(20.6, 1.5)node[right]{$\mathcal{A}$};
\draw[->](1.5, -5)--(1.75, -4.5);
\draw(1.75, -4.5)--(2, -4);
\draw[->](3, -3.5)--(2.5, -3.75);
\draw(2.5, -3.75)--(2, -4);
\draw[->](2, -4)--(1.5, -3.5);
\draw(1.5, -3.5)--(1, -3);
\draw[->](1, -3)--(0.5, -3.25);
\draw(0.5, -3.25)--(0, -3.5);
\draw[->](1, -3)--(1.25, -2.5);
\draw(1.25, -2.5)--(1.5, -2);
\fill(2, -4)circle(0.06);
\fill(1, -3)circle(0.06);
\draw(2, -4)node[below right]{$m$};
\draw(1, -3)node[above left]{$m^{\vee}$};
\draw(1.5, -2)node[above]{$\mathcal{A}$};
\draw(0, -3.5)node[left]{$\mathcal{A}$};
\draw(1.5, -5)node[below]{$\mathcal{A}$};
\draw(3, -3.5)node[right]{$\mathcal{A}$};
\draw(1.5, -3.25)node[right]{$\mathcal{A}$};
\draw(4.5, -3.5)node{$=$};
\draw[->](7.5, -5)--(7.25, -4.5);
\draw(7.25, -4.5)--(7, -4);
\draw[->](7, -4)--(6.5, -3.75);
\draw(6.5, -3.75)--(6, -3.5);
\draw[->](7, -4)--(7.5, -3.5);
\draw(7.5, -3.5)--(8, -3);
\draw[->](8, -3)--(7.75, -2.5);
\draw(7.75, -2.5)--(7.5, -2);
\draw[->](9, -3.5)--(8.5, -3.25);
\draw(8.5, -3.25)--(8, -3);
\fill(7, -4)circle(0.06);
\fill(8, -3)circle(0.06);
\draw(7, -4)node[below left]{$m^{\vee}$};
\draw(8, -3)node[above right]{$m$};
\draw(7.5, -2)node[above]{$\mathcal{A}$};
\draw(6, -3.5)node[left]{$\mathcal{A}$};
\draw(7.5, -5)node[below]{$\mathcal{A}$};
\draw(9, -3.5)node[right]{$\mathcal{A}$};
\draw(7.5, -3.75)node[right]{$\mathcal{A}$};
\draw[->](13, -5)--(13, -4.5);
\draw(13, -4.5)--(13, -4);
\draw[->](13, -3)--(13, -2.5);
\draw(13, -2.5)--(13, -2);
\draw[->](13, -4) arc (270: 180: 0.5);
\draw(12.5, -3.5) arc (180: 90: 0.5);
\draw[->](13, -4) arc (270: 360: 0.5);
\draw(13.5, -3.5) arc (0: 90: 0.5);
\fill(13, -4)circle(0.06);
\fill(13, -3)circle(0.06);
\draw(13, -4)node[below right]{$m^{\vee}$};
\draw(13, -3)node[above right]{$m$};
\draw(13, -5)node[below]{$\mathcal{A}$};
\draw(13, -2)node[above]{$\mathcal{A}$};
\draw(12, -3.5)node{$\mathcal{A}$};
\draw(14, -3.5)node{$\mathcal{A}$};
\draw(15, -3.5)node{$=$};
\draw[->](16.5, -5)--(16.5, -3.5);
\draw(16.5, -3.5)--(16.5, -2);
\draw(16.5, -5)node[below]{$\mathcal{A}$};
\draw(16.5, -2)node[above]{$\mathcal{A}$};
\end{tikzpicture}
\caption{The conditions on $\mathcal{A}$ and $m(m^{\vee})$. We require associativity which corresponds to an anomaly-free condition in a finite group symmetry. The bottom right figure represents the invariance under the change of the triangulation.}
\label{Amcondition}
\end{figure}

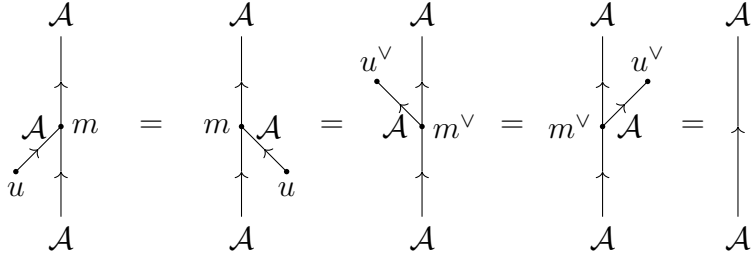
\begin{figure}[ht]
\centering
\begin{tikzpicture}[scale = 0.6]
\draw[->](0, 1)--(0.5, 1.5);
\draw(0.5, 1.5)--(1, 2);
\draw[->](1, 0)--(1, 1);
\draw[->](1, 1)--(1, 3);
\draw(1, 3)--(1, 4);
\fill(0, 1)circle(0.06);
\fill(1, 2)circle(0.06);
\draw(1, 2)node[right]{$m$};
\draw(0, 1)node[below]{$u$};
\draw(1, 0)node[below]{$\mathcal{A}$};
\draw(1, 4)node[above]{$\mathcal{A}$};
\draw(0.4, 1.5)node[above]{$\mathcal{A}$};
\draw(3, 2)node{$=$};
\draw[->](6, 1)--(5.5, 1.5);
\draw(5.5, 1.5)--(5, 2);
\draw[->](5, 0)--(5, 1);
\draw[->](5, 1)--(5, 3);
\draw(5, 3)--(5, 4);
\fill(6, 1)circle(0.06);
\fill(5, 2)circle(0.06);
\draw(5, 2)node[left]{$m$};
\draw(6, 1)node[below]{$u$};
\draw(5, 0)node[below]{$\mathcal{A}$};
\draw(5, 4)node[above]{$\mathcal{A}$};
\draw(5.6, 1.5)node[above]{$\mathcal{A}$};
\draw(7, 2)node{$=$};
\draw[->](9, 2)--(8.5, 2.5);
\draw(8.5, 2.5)--(8, 3);
\draw[->](9, 0)--(9, 1);
\draw[->](9, 1)--(9, 3);
\draw(9, 3)--(9, 4);
\fill(8, 3)circle(0.06);
\fill(9, 2)circle(0.06);
\draw(9, 2)node[right]{$m^{\vee}$};
\draw(8, 3)node[above]{$u^{\vee}$};
\draw(9, 0)node[below]{$\mathcal{A}$};
\draw(9, 4)node[above]{$\mathcal{A}$};
\draw(8.4, 2.5)node[below]{$\mathcal{A}$};
\draw(11, 2)node{$=$};
\draw[->](13, 2)--(13.5, 2.5);
\draw(13.5, 2.5)--(14, 3);
\draw[->](13, 0)--(13, 1);
\draw[->](13, 1)--(13, 3);
\draw(13, 3)--(13, 4);
\fill(14, 3)circle(0.06);
\fill(13, 2)circle(0.06);
\draw(13, 2)node[left]{$m^{\vee}$};
\draw(14, 3)node[above]{$u^{\vee}$};
\draw(13, 0)node[below]{$\mathcal{A}$};
\draw(13, 4)node[above]{$\mathcal{A}$};
\draw(13.6, 2.5)node[below]{$\mathcal{A}$};
\draw(15, 2)node{$=$};
\draw[->](16, 0)--(16, 2);
\draw(16, 2)--(16, 4);
\draw(16, 0)node[below]{$\mathcal{A}$};
\draw(16, 4)node[above]{$\mathcal{A}$};
\end{tikzpicture}
\caption{The conditions on $\mathcal{A}$, $m(m^{\vee})$ and $u(u^{\vee})$.}
\label{Amucondition}
\end{figure}

Now we generalize this idea to a fusion category $\mathcal{C}$. For this purpose, we prepare the following data. The first is a TDL $\mathcal{A}$ which is non-simple in general. The second is a multiplication morphism $m \in \mathrm{Hom}(\mathcal{A} \times \mathcal{A}, \mathcal{A})$ and its adjoint $m^{\vee} \in \mathrm{Hom}(\mathcal{A}, \mathcal{A} \times \mathcal{A})$. We require the conditions as in Figure \ref{Amcondition}.
The upper diagrams represent the associativity conditions and correspond to the anomaly-free conditions of gauging a finite group $G$. The lower diagrams remove the ambiguity of the triangulation of the spacetime \cite{Chung:1993xr}. The third is a unit $u \in \mathrm{Hom}(1, \mathcal{A})$ and co-unit $u^{\vee} \in \mathrm{Hom}(\mathcal{A}, 1)$ which ensure that vacuum survives under gauging. We also require the conditions as in Figure \ref{Amucondition}. 
A set $(\mathcal{A}, m, m^{\vee}, u, u^{\vee})$ which satisfies the above conditions are called a symmetric separable special Frobenius algebra \cite{Bhardwaj:2017xup, Ostrik:2001xnt, Fuchs:2002cm}. We will simply denote it as $\mathcal{A}$ or $(\mathcal{A}, m)$ when there is no confusion.

Given a symmetric separable special Frobenius algebra $\mathcal{A}$, we define the gauged theory $\mathcal{T}/\mathcal{A}$ by inserting a fine-enough mesh of $\mathcal{A}$\footnote{An alternative description based on the SymTFT can be found in \cite{Chen:2024ulc,Putrov:2024uor}.} 
\begin{align}
     Z_{\mathcal{T}/\mathcal{A}} =&\ \begin{tikzpicture}[scale=0.6,baseline={([yshift=-.5ex]current bounding box.center)},vertex/.style={anchor=base,
    circle,fill=black!25,minimum size=18pt,inner sep=2pt},scale=0.50]        
    \draw(0,0)--(6, 0)--(6,6)--(0,6)--(0,0);
    \draw[->](3,0)--(3.5,1);
    \draw(3.5,1)--(4,2);
    \draw[->](6,3)--(5,2.5);
    \draw(5,2.5)--(4,2);
    \draw[->](4,2)--(3,3);
    \draw(3,3)--(2,4);
    \draw[->](2,4)--(1,3.5);
    \draw(1,3.5)--(0,3);
    \draw[->](2,4)--(2.5,5);
    \draw(2.5,5)--(3,6);
    \draw(2.5,5)node[right]{$\mathcal{A}$};
    \draw(5,2.5)node[above]{$\mathcal{A}$};
    \draw(2.3,2.3)node{$\mathcal{A}$};
    \fill(4,2)circle(0.1);
    \fill(2,4)circle(0.1);
    \draw(3.8,2.2)node[below right]{$m$};
    \draw(2.8,3.8)node[above left]{$m^{\vee}$};
    \end{tikzpicture} =  \sum_{\substack{L_x, L_y, L_z \in \mathcal{A}\\ L_z \in L_x \times L_y}} m_{xy}^{z}(m^{\vee})_{yx}^{z} \begin{tikzpicture}[scale=0.6,baseline={([yshift=-.5ex]current bounding box.center)},vertex/.style={anchor=base,
    circle,fill=black!25,minimum size=18pt,inner sep=2pt},scale=0.50]        
    \draw(0,0)--(6, 0)--(6,6)--(0,6)--(0,0);
    \draw[red,->](3,0)--(3.5,1);
    \draw[red](3.5,1)--(4,2);
    \draw[blue,->](6,3)--(5,2.5);
    \draw[blue](5,2.5)--(4,2);
    \draw[green,->](4,2)--(3,3);
    \draw[green](3,3)--(2,4);
    \draw[blue,->](2,4)--(1,3.5);
    \draw[blue](1,3.5)--(0,3);
    \draw[red,->](2,4)--(2.5,5);
    \draw[red](2.5,5)--(3,6);
    \draw[red](2.5,5)node[right]{$L_x$};
    \draw[blue](5,2.5)node[above]{$L_y$};
    \draw[green](2.3,2.3)node{$L_z$};
    \end{tikzpicture}.
\end{align}
The last diagram is defined by
\begin{align}
    Z_{L_xL_y}^{L_z}(\tau) := \begin{tikzpicture}[scale=0.6,baseline={([yshift=-.5ex]current bounding box.center)},vertex/.style={anchor=base,
    circle,fill=black!25,minimum size=18pt,inner sep=2pt},scale=0.50]        
    \draw(0,0)--(6, 0)--(6,6)--(0,6)--(0,0);
    \draw[red,->](3,0)--(3.5,1);
    \draw[red](3.5,1)--(4,2);
    \draw[blue,->](6,3)--(5,2.5);
    \draw[blue](5,2.5)--(4,2);
    \draw[green,->](4,2)--(3,3);
    \draw[green](3,3)--(2,4);
    \draw[blue,->](2,4)--(1,3.5);
    \draw[blue](1,3.5)--(0,3);
    \draw[red,->](2,4)--(2.5,5);
    \draw[red](2.5,5)--(3,6);
    \draw[red](2.5,5)node[right]{$L_x$};
    \draw[blue](5,2.5)node[above]{$L_y$};
    \draw[green](2.3,2.3)node{$L_z$};
    \end{tikzpicture} = \sum_{(i, \bar{i}) \in \mathcal{H}_{1|x}^{\mathrm{prim}}} (\hat{L}_y)_{L_z}^{i, \bar{i}}\chi_{i}(\tau)\bar{\chi}_{\bar{i}}(\bar{\tau})
\end{align}
where the number $(\hat{L}_y)_{L_z}^{i, \bar{i}}$ represents the action of $L_y$ on the primary $(i, \bar{i})$ of $\mathcal{H}_{1|x}$ under the insertion of $L_z$. $(\hat{L}_y)_{L_z}^{i, \bar{i}}$ must be compatible with the constraints derived from the modular S-transformation and F-move. By performing the modular S-transformation,
\begin{align}
    Z_{L_xL_y}^{L_z}\left(-\dfrac{1}{\tau}\right) = \sum_{(i, \bar{i}) \in \mathcal{H}_{1|x}^{\mathrm{prim}}}\sum_{j, \bar{j}} (\hat{L}_y)_{L_z}^{i, \bar{i}}S_{ij}S_{\bar{i}\bar{j}^{*}}\chi_{j}(\tau)\bar{\chi}_{\bar{j}}(\bar{\tau}).\label{sxyz}
\end{align}
Meanwhile, by F-move,
\begin{align}
    Z_{L_xL_y}^{L_z}\left(-\dfrac{1}{\tau}\right) =&\ \begin{tikzpicture}[scale=0.6,baseline={([yshift=-.5ex]current bounding box.center)},vertex/.style={anchor=base,
    circle,fill=black!25,minimum size=18pt,inner sep=2pt},scale=0.50]        
    \draw(0,0)--(6, 0)--(6,6)--(0,6)--(0,0);
    \draw[blue,->](3,0)--(2.5,1);
    \draw[blue](2.5,1)--(2,2);
    \draw[red,->](0,3)--(1,2.5);
    \draw[red](2,2)--(1,2.5);
    \draw[green][->](2,2)--(3,3);
    \draw[green](3,3)--(4,4);
    \draw[blue][->](4,4)--(3.5,5);
    \draw[blue](3.5,5)--(3,6);
    \draw[red](6,3)--(5,3.5);
    \draw[red,<-](5,3.5)--(4,4);
    \draw[red](1,2.3)node[above]{$L_x$};
    \draw[blue](3.8,5)node[left]{$L_y$};
    \draw[green](3.7,2.3)node{$L_z$};
    \end{tikzpicture} = \sum_{L_w} F_{L_xL_y\overline{L_x}}^{L_y}(L_z, L_w)\ \begin{tikzpicture}[scale=0.6,baseline={([yshift=-.5ex]current bounding box.center)},vertex/.style={anchor=base,
    circle,fill=black!25,minimum size=18pt,inner sep=2pt},scale=0.50]        
    \draw(0,0)--(6, 0)--(6,6)--(0,6)--(0,0);
    \draw[blue,->](3,0)--(3.5,1);
    \draw[blue](3.5,1)--(4,2);
    \draw[red,](6,3)--(5,2.5);
    \draw[red,<-](5,2.5)--(4,2);
    \draw[orange,->](4,2)--(3,3);
    \draw[orange](3,3)--(2,4);
    \draw[red](2,4)--(1,3.5);
    \draw[red,<-](1,3.5)--(0,3);
    \draw[blue,->](2,4)--(2.5,5);
    \draw[blue](2.5,5)--(3,6);
    \draw[blue](2.5,5)node[right]{$L_y$};
    \draw[red](5,2.5)node[above]{$L_x$};
    \draw[orange](2.3,2.3)node{$L_w$};
    \end{tikzpicture}\notag\\
    =& \sum_{L_w}\sum_{(k, \bar{k}) \in \mathcal{H}_{1|y}^{\mathrm{prim}}} F_{L_xL_y\overline{L_x}}^{L_y}(L_z, L_w)(\hat{\overline{L_x}})_{L_w}^{k, \bar{k}}\chi_{k}(\tau)\bar{\chi}_{\bar{k}}(\bar{\tau}).\label{fxyz}
\end{align}
Comparing (\ref{sxyz}) and (\ref{fxyz}), we obtain
\begin{align}
    \sum_{(i, \bar{i}) \in \mathcal{H}_{1|x}^{\mathrm{prim}}} (\hat{L}_y)_{L_z}^{i, \bar{i}}S_{ij}S_{\bar{i}\bar{j}^{*}} = \begin{cases}
        \displaystyle\sum_{L_w} F_{L_xL_y\overline{L_x}}^{L_y}(L_z, L_w)(\hat{\overline{L_x}})_{L_w}^{j, \bar{j}}\ \ \ \mathrm{if}\ (j, \bar{j}) \in \mathcal{H}_{1|y}^{\mathrm{prim}}\\
        0\ \ \ \ \ \ \ \ \ \ \ \ \ \ \ \ \ \ \ \ \ \ \ \ \ \ \ \ \ \ \ \ \ \ \ \ \mathrm{otherwise}.
    \end{cases}\label{constraintsf}
\end{align}

At this point, it is useful to mention the relation between the algebra objects and the interface. In general, we denote $I_{\mathcal{T}_1|\mathcal{T}_2}$ as the interface between two theories $\mathcal{T}_1$ and $\mathcal{T}_2$. It is argued that when $I_{\mathcal{T}_1|\mathcal{T}_2}$ is topological, $\mathcal{T}_1$ and $\mathcal{T}_2$ are related by gauging an algebra $\mathcal{A}$ \cite{Diatlyk:2023fwf}. Here, our goal is to relate $\mathcal{A}$ with the interface $I_{\mathcal{T}|\mathcal{T}/\mathcal{A}}$.

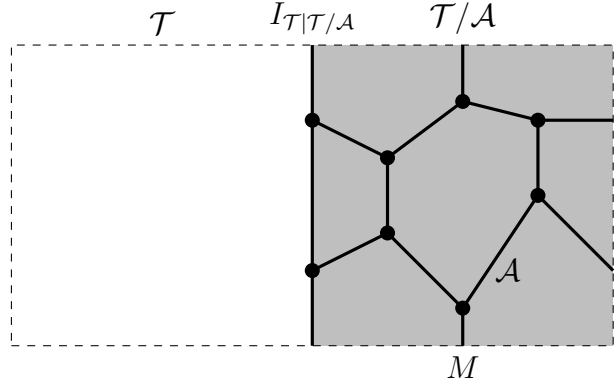
\begin{figure}[ht]
\centering
\begin{tikzpicture}
\fill[lightgray](0, 0)--(4, 0)--(4, 4)--(0, 4)--(0, 0)--cycle;
\draw[very thick](0, 0)--(0,4);
\draw(-2, 4)node[above]{$\mathcal{T}$};
\draw(2, 4)node[above]{$\mathcal{T}/\mathcal{A}$};
\draw(0, 4)node[above]{$I_{\mathcal{T}|\mathcal{T}/\mathcal{A}}$};
\draw(2, 0)node[below]{$M$};
\draw[very thick](0, 1)--(1, 1.5);
\draw[very thick](0, 3)--(1, 2.5);
\draw[very thick](1, 1.5)--(1, 2.5);
\draw[very thick](1, 1.5)--(2, 0.5);
\draw[very thick](1, 2.5)--(2, 3.25);
\draw[very thick](2, 0.5)--(3, 2);
\draw[very thick](2, 3.25)--(3, 3);
\draw[very thick](3, 2)--(3, 3);
\draw[very thick](3, 2)--(4, 1);
\draw[very thick](3, 3)--(4, 3);
\draw[very thick](2, 3.25)--(2, 4);
\draw[very thick](2, 0)--(2, 0.5);
\draw(2.6, 1)node{$\mathcal{A}$};
\fill(1, 1.5)circle(0.1);
\fill(1, 2.5)circle(0.1);
\fill(2, 0.5)circle(0.1);
\fill(2, 3.25)circle(0.1);
\fill(3, 2)circle(0.1);
\fill(3, 3)circle(0.1);
\fill(0, 1)circle(0.1);
\fill(0, 3)circle(0.1);
\draw[dashed](-4, 0)--(4, 0)--(4, 4)--(-4, 4)--(-4, 0);
\end{tikzpicture}
\caption{The topological interface between $\mathcal{T}$ and $\mathcal{T}/\mathcal{A}$ from half-space gauging.}
\label{interface}
\end{figure}

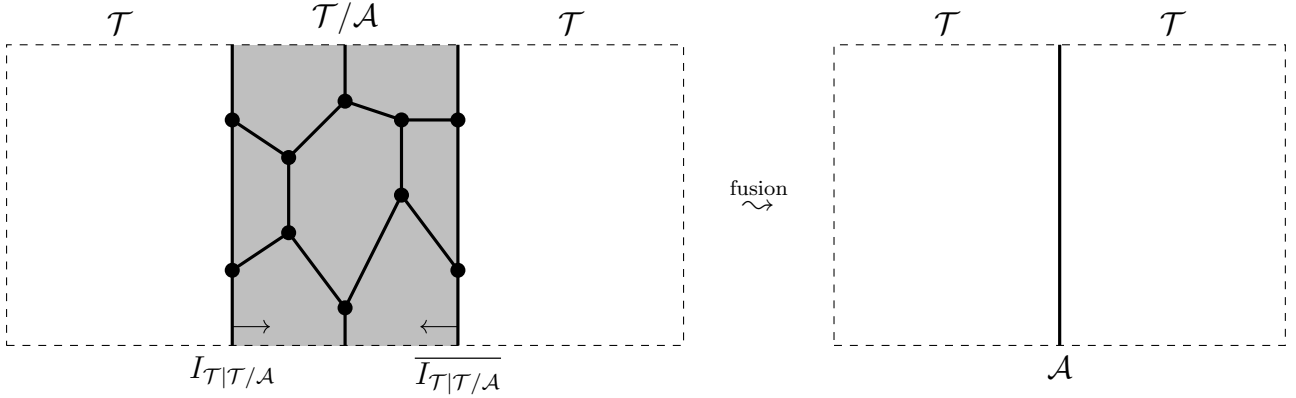
\begin{figure}[ht]
\centering
\begin{tikzpicture}
\fill[lightgray](0, 0)--(3, 0)--(3, 4)--(0, 4)--(0, 0)--cycle;
\draw[very thick](0, 0)--(0, 4);
\draw[very thick](3, 0)--(3, 4);
\draw(7, 2)node{$\stackrel{\mathrm{fusion}}{\leadsto}$};
\draw[very thick](11, 0)--(11, 4);
\draw(-1.5, 4)node[above]{$\mathcal{T}$};
\draw(1.5, 4)node[above]{$\mathcal{T}/\mathcal{A}$};
\draw(4.5, 4)node[above]{$\mathcal{T}$};
\draw(9.5, 4)node[above]{$\mathcal{T}$};
\draw(12.5, 4)node[above]{$\mathcal{T}$};
\draw(0, 0)node[below]{$I_{\mathcal{T}|\mathcal{T}/\mathcal{A}}$};
\draw(3, 0)node[below]{$\overline{I_{\mathcal{T}|\mathcal{T}/\mathcal{A}}}$};
\draw(11, 0)node[below]{$\mathcal{A}$};
\draw[->](0, 0.25)--(0.5, 0.25);
\draw[->](3, 0.25)--(2.5, 0.25);
\draw[very thick](0, 1)--(0.75, 1.5);
\draw[very thick](0, 3)--(0.75, 2.5);
\draw[very thick](0.75, 1.5)--(0.75, 2.5);
\draw[very thick](0.75, 1.5)--(1.5, 0.5);
\draw[very thick](0.75, 2.5)--(1.5, 3.25);
\draw[very thick](1.5, 0.5)--(2.25, 2);
\draw[very thick](1.5, 3.25)--(2.25, 3);
\draw[very thick](2.25, 2)--(2.25, 3);
\draw[very thick](2.25, 2)--(3, 1);
\draw[very thick](2.25, 3)--(3, 3);
\draw[very thick](1.5, 3.25)--(1.5, 4);
\draw[very thick](1.5, 0)--(1.5, 0.5);
\fill(0.75, 1.5)circle(0.1);
\fill(0.75, 2.5)circle(0.1);
\fill(1.5, 0.5)circle(0.1);
\fill(1.5, 3.25)circle(0.1);
\fill(2.25, 2)circle(0.1);
\fill(2.25, 3)circle(0.1);
\fill(0, 1)circle(0.1);
\fill(0, 3)circle(0.1);
\fill(3, 3)circle(0.1);
\fill(3, 1)circle(0.1);
\draw[dashed](-3, 0)--(6, 0)--(6, 4)--(-3, 4)--(-3, 0);
\draw[dashed](8, 0)--(14, 0)--(14, 4)--(8, 4)--(8, 0);
\end{tikzpicture}
\caption{The fusion of the topological interface $I_{\mathcal{T}|\mathcal{T}/\mathcal{A}}$ and its orientation reversal $\overline{I_{\mathcal{T}|\mathcal{T}/\mathcal{A}}}$ gives the algebra object $\mathcal{A}$.}
\label{interfacefusion}
\end{figure}

Consider a theory $\mathcal{T}$ with a symmetry $\mathcal{C}$ and pick an algebra $(\mathcal{A}, m)$. For this purpose, instead of gauging $(\mathcal{A}, m)$ on the whole spacetime, we gauge on the half-spacetime $M$. In doing so, we impose the canonical topological boundary condition on $\partial M$ by inserting $\mathcal{A}$ itself along the boundary $\partial M$. This defines the topological interface $I_{\mathcal{T}|\mathcal{T}/\mathcal{A}}$ on $\partial M$ as in Figure \ref{interface}.
Next, we split the spacetime into three regions and gauge $\mathcal{A}$ only on the middle region of the spacetime. When we narrow its width, the gauged region in which a mesh of $\mathcal{A}$ inserted becomes smaller. The effect of the gauging, then eventually reduces to the configuration where only $\mathcal{A}$ is inserted on the fused boundary $\partial M$ as in Figure \ref{interfacefusion}. 
It means that we can obtain the algebra object $\mathcal{A}$ by fusing two interfaces 
\begin{align}
    \mathcal{A} = I_{\mathcal{T}|\mathcal{T}/\mathcal{A}} \times I_{\mathcal{T}/\mathcal{A}|\mathcal{T}} = I_{\mathcal{T}|\mathcal{T}/\mathcal{A}} \times \overline{I_{\mathcal{T}|\mathcal{T}/\mathcal{A}}}.\label{algebrainterface}
\end{align}
Conversely, an algebra object $\mathcal{A}^{*}$ which implements the inverse gauging is
\begin{align}
    \mathcal{A}^{*} =  I_{\mathcal{T}/\mathcal{A}|\mathcal{T}} \times I_{\mathcal{T}|\mathcal{T}/\mathcal{A}} = \overline{I_{\mathcal{T}|\mathcal{T}/\mathcal{A}}} \times I_{\mathcal{T}|\mathcal{T}/\mathcal{A}}.\label{Adual}
\end{align}

Let us show two useful applications of this relation \cite{Diatlyk:2023fwf}. The first one is ``the quantum dimensions of the $\mathcal{A}$ and $\mathcal{A}^*$ related by the inverse gauging are the same". We can show this by assuming the interface $I_{\mathcal{T}_1|\mathcal{T}_2}$ maps the vacuum of $\mathcal{T}_1$ to the vacuum of $\mathcal{T}_2$, and taking the vacuum expectation value of equation (\ref{Adual}). It gives
\begin{align}
    _{\mathcal{T}/\mathcal{A}}\bra{0}\mathcal{A}^{*}\ket{0}_{\mathcal{T}/\mathcal{A}} &= _{\mathcal{T}/\mathcal{A}}\bra{0}\overline{I_{\mathcal{T}|\mathcal{T}/\mathcal{A}}}\ket{0}_{\mathcal{T}}\ _{\mathcal{T}}\bra{0}I_{\mathcal{T}|\mathcal{T}/\mathcal{A}}\ket{0}_{\mathcal{T}/\mathcal{A}}\notag\\
    &= _{\mathcal{T}}\bra{0}\mathcal{A}\ket{0}_{\mathcal{T}}.\label{AAdualqdim}
\end{align}
Here, the right-hand-side object, the vacuum expectation value of $\mathcal{A}$, is called the quantum dimension of $\mathcal{A}$.

The second one is ``in the sequential gauging, the quantum dimension of the total algebra is given by the product of those of the one-step algebra". To show this, we consider the sequential gauging $\mathcal{T} \to \mathcal{T}^{\prime} = \mathcal{T}/\mathcal{A} \to \mathcal{T}^{\prime\prime} = \mathcal{T}^{\prime}/\mathcal{A}^{\prime}$. The fusion $I_{\mathcal{T}|\mathcal{T}^{\prime\prime}} = I_{\mathcal{T}|\mathcal{T}^{\prime}} \times I_{\mathcal{T}^{\prime}|\mathcal{T}^{\prime\prime}}$ is a topological interface between $\mathcal{T}$ and $\mathcal{T}^{\prime\prime}$ and an algebra $\mathcal{A}^{\prime\prime}$ which implements the gauging $\mathcal{T} \to \mathcal{T}^{\prime\prime}$ in one step is 
\begin{align}
    \mathcal{A}^{\prime\prime} = I_{\mathcal{T}|\mathcal{T}^{\prime\prime}} \times \overline{I_{\mathcal{T}|\mathcal{T}^{\prime\prime}}} = I_{\mathcal{T}|\mathcal{T}^{\prime}} \times I_{\mathcal{T}^{\prime}|\mathcal{T}^{\prime\prime}} \times \overline{I_{\mathcal{T}^{\prime}|\mathcal{T}^{\prime\prime}}} \times \overline{I_{\mathcal{T}|\mathcal{T}^{\prime}}}.\label{algebraonestep}
\end{align}
In the same way as (\ref{AAdualqdim}), taking the vacuum expectation value of (\ref{algebraonestep}) gives
\begin{align}
    _{\mathcal{T}}\bra{0}\mathcal{A}^{\prime\prime}\ket{0}_{\mathcal{T}} = _{\mathcal{T}}\bra{0}\mathcal{A}\ket{0}_{\mathcal{T}}\ _{\mathcal{T}^{\prime}}\bra{0}\mathcal{A}^{\prime}\ket{0}_{\mathcal{T}^{\prime}}\label{qdimalgbraonestep} \ .
\end{align}

These two relations are useful in finding  {\it one} algebra that implements the desired gauging. We, however, should note that it is possible that $\mathcal{T}/\mathcal{A}$ and $\mathcal{T}/\mathcal{A'}$ are the same even though the quantum dimension of $\mathcal{A}$ and $\mathcal{A}'$ is different, and hence there can exist algebra that gives the same theory even without satisfying these relations.
We may be able to find a simpler algebra that results in the same gauged theory as we will see.\footnote{One such example is gauging the Fibonacci algebra $\mathcal{A} = 1+W (= W \times W)$, whose quantum dimension is not one, but it results in trivial gauging: $\mathcal{T}/\mathcal{A} = \mathcal{T}$.}

In (\ref{algebrainterface}), we focused on only the algebra object. Let us now briefly mention the algebra structure such as the multiplication morphism \cite{EGNO}. Because the TDLs of $\mathcal{T}$, which are the objects of $\mathcal{C}$, act on the topological interface $I_{\mathcal{T}|\mathcal{T}/\mathcal{A}}$ from the left, the topological interfaces between $\mathcal{T}$ and $\mathcal{T}/\mathcal{A}$ form the left $\mathcal{C}$-module category $\mathrm{Mod}_{\mathcal{C}}(\mathcal{A})$. For $I_1, I_2 \in \mathrm{Mod}_{\mathcal{C}}(\mathcal{A})$, we can define the object $\underline{\mathrm{Hom}}(I_1, I_2) \in \mathcal{C}$ which satisfies the property
\begin{align}
    \mathrm{Hom}_{\mathrm{Mod}_{\mathcal{C}}(\mathcal{A})}(L \times I_1, I_2) \cong \mathrm{Hom}_{\mathcal{C}}(L, \underline{\mathrm{Hom}}(I_1, I_2))
\end{align}
for any $L \in \mathcal{C}$. This object is called the internal Hom from $I_1$ to $I_2$.

By using this property, the canonical morphism
\begin{align}
    m_{I_1I_2I_3} : \underline{\mathrm{Hom}}(I_2, I_3) \times \underline{\mathrm{Hom}}(I_1, I_2) \to \underline{\mathrm{Hom}}(I_1, I_3)
\end{align}
can be defined. In particular, $m_{III}$ is the canonical multiplication morphism of $\underline{\mathrm{Hom}}(I, I)$, which is what we have denoted as $I \times \overline{I}$. With this canonical multiplication morphism, $\underline{\mathrm{Hom}}(I, I)$ admits the algebra structure. The way to get an algebra object described in Figure \ref{interfacefusion} is the physical interpretation of the internal Hom construction.\footnote{The internal Hom construction for the regular module category appeared in relation to the weakly symmetric boundary condition studied in \cite{Choi:2023xjw}. Note that our module category based on the interface is not always regular, so we can construct non-trivial gauging.} In practice, however, performing the internal Hom construction explicitly can be cumbersome, so we will take an alternative way to determine the multiplication morphism from the consistency conditions for the algebra.

\subsection{Exchanging \texorpdfstring{$\mathcal{M}(A_{10}, D_7)$}{TEXT} and \texorpdfstring{$\mathcal{M}(A_{10}, E_6)$}{TEXT}}
We will now show the main claim of this paper ``what's done can be undone": We find the algebra that exchanges {$\mathcal{M}(A_{10}, D_7)$} and $\mathcal{M}(A_{10}, E_6)${\footnote{We can generalize the following discussion to the exchange of $\mathcal{M}(A_{h - 1}, D_7)$ and $\mathcal{M}(A_{h - 1}, E_6)$. However, we need to use the different multiplication morphism depending on $h$ since the quantum dimension of the algebra object depends on $h$. See (\ref{qd122}).}} by performing the explicit gauging.  

Our starting point is three algebra objects of $\mathcal{M}(A_{10}, A_{11})$
\begin{align}
    \mathcal{A}_{A_{11}} = 1 + L_{(1, 3)},\ \mathcal{A}_{D_{7}} = 1 + L_{(1, 11)},\ \mathcal{A}_{E_6} = 1 + L_{(1, 7)}
\end{align}
with the unique multiplication morphisms \cite{Bhardwaj:2017xup, Diatlyk:2023fwf, Ostrik:2001xnt, Kirillov:2001ti}. According to Lemma 8 in \cite{Ostrik:2001xnt} ``Let $\mathcal{C}$ be a rigid monoidal category and $X$ be an irreducible object of $\mathcal{C}$. Assume that $\mathrm{Hom}(X\times X,X)$ is one dimensional. Then $A = 1+X$ has at most one structure of a semisimple algebra in $\mathcal{C}$'', $\mathcal{A}_{A_{11}}$ and $\mathcal{A}_{E_{6}}$ have the unique algebra structure because $\mathrm{dim}\ \mathrm{Hom}(L_{(1, 3)} \times L_{(1, 3)}, L_{(1, 3)}) = \mathrm{dim}\ \mathrm{Hom}(L_{(1, 7)} \times L_{(1, 7)}, L_{(1, 7)}) = 1$. From the fusion rule $L_{(1, 11)}^2 = 1$, $L_{(1, 11)}$ is the $\mathbb{Z}_2$ line. Because $H^2(\mathbb{Z}_2, U(1))$ is trivial, there is no possible non-trivial discrete torsion here, and $\mathcal{A}_{D_{7}}$ has the unique multiplication morphism
\begin{align}
    m_{11}^{1} = m_{1\ L_{(1, 11)}}^{L_{(1, 11)}} = m_{L_{(1, 11)}\ 1}^{L_{(1, 11)}} = m_{L_{(1, 11)}\ L_{(1, 11)}}^{1} = \dfrac{1}{\sqrt{2}}.
\end{align}

Because $\mathcal{A}_{A_{11}} = L_{(1, 2)} \times L_{(1, 2)}$, the first algebra is Morita trivial and implements trivial gauging \cite{Diatlyk:2023fwf, Roumpedakis:2022aik, Choi:2023xjw}. Since $L_{(1, 11)}$ is non-anomalous $\mathbb{Z}_2$ invertible symmetry line, the second algebra implements $\mathbb{Z}_2$ orbifold and results in $\mathcal{M}(A_{10}, D_7)$. As known in the orbifold construction, the exchanged theory $\mathcal{M}(A_{10}, D_7)$ has the dual $\mathbb{Z}_2$ symmetry and returns to $\mathcal{M}(A_{10}, A_{11})$ when we gauge this symmetry. 

The most interesting gauging for us is based on the third algebra.
In \cite{Ostrik:2001xnt, Kirillov:2001ti}, it was shown that the third algebra implements the non-invertible symmetry gauging from $\mathcal{M}(A_{10}, A_{11})$ to $\mathcal{M}(A_{10}, E_6)$. In \cite{Diatlyk:2023fwf}, it is claimed that the dual algebra object $\mathcal{A}_{E_6}^{*}$ which satisfies the constraint (\ref{AAdualqdim}) on the quantum dimension  is
\begin{align}
    \mathcal{A}_{E_6}^{*} = 1 + L_{(1, 2, 2)}.
\end{align}

Let us now consider the sequential gauging of $\mathcal{M}(A_{10}, E_6)$ to $\mathcal{M}(A_{10}, A_{11})$ and to $\mathcal{M}(A_{10}, D_7)$. From the formula (\ref{qdimalgbraonestep}), the quantum dimensions of {\it one} algebra that implements the sequential gauging from $\mathcal{M}(A_{10},E_{6})$ to $\mathcal{M}(A_{10}, D_{7})$ at a time must be $2(3 + \sqrt{3})$ because their quantum dimensions are $\braket{\mathcal{A}_{E_6}} = \braket{\mathcal{A}_{E_6}^{*}} = 3 + \sqrt{3}$\footnote{As we will see in (\ref{122zu2}), $\braket{L_{(1, 2, 2)}} = 2 + \sqrt{3}$. The fusion rule itself gives a severe constraint on the quantum dimensions. In subsection \ref{E6TDL} we showed $\mathcal{M}(A_{h - 1}, E_{6})$ has TDLs listed in (\ref{E6TDLs}) obeying the fusion rules (\ref{E6fusion}). By taking the expectation values of the fusion rule, we see that $\braket{L_{(1, 2, 2)}} = 2 \pm \sqrt{3}$, which is consistent with the explicit computation in (\ref{122zu2}), but is not determined uniquely. In general, \begin{align}
    \braket{L_{(1, 2, 2)}} = 
    \begin{cases}
         2 + \sqrt{3}\ \ \ \ h = \pm 1\ (\mathrm{mod}\ 12)\\
         2 - \sqrt{3}\ \ \ \ h = \pm 5\ (\mathrm{mod}\ 12).
    \end{cases}\label{qd122}
\end{align}
The quantum dimension of $\braket{L_{(1, 4, 2)}}$ is also given by the same formula.} and $\braket{\mathcal{A}_{D_7}} =2$. The most natural algebra object $\mathcal{\mathcal{A}}_1$ which satisfies the constraint on the quantum dimension is 
\begin{align}
    \mathcal{A}_1 = (1 + \eta)(1 + L_{(1, 4, 2)})
\end{align}
with a particular multiplication morphism induced by the sequential gauging (i.e. induced by the internal $\mathrm{Hom}$) discussed in the previous subsection. Since demonstrating this gauging explicitly should be challenging due to the large number of diagrams to be considered, we look for a simpler algebra that implements the same gauging.

One important observation here is that, as we have seen in the \ref{E6TDL}, because $\mathcal{M}(A_{10}, E_{6})$ has $\mathbb{Z}_2$ Tambara-Yamagami category, we can rewrite the algebra object as 
\begin{align}
    \mathcal{A}_1 = \mathcal{N}(1 + L_{(1, 4, 2)})\overline{\mathcal{N}} \ .
\end{align}
{\it If} the multiplication morphism is consistent with this factorization, the algebra $\mathcal{A}_1$ is Morita equivalent to the algebra $\mathcal{A}_2 = 1 + L_{(1, 4, 2)}$, and gauging $\mathcal{A}_1$ is equivalent to gauging $\mathcal{A}_2$.\footnote{The importance of the choice of the multiplication morphisms here can be understood by the following observation. As an algebra object, $\mathcal{A}_1$ can also be rewritten as $\mathcal{A}' = (1 + \eta)(1 + L_{(1, 2, 2)}) = \mathcal{N}(1 + L_{(1, 2, 2)})\overline{\mathcal{N}}$, and one might wonder if it is also Morita equivalent to $1+ L_{(1,2,2)} = \mathcal{A}^*_{E_6}$. Whether the algebra (i.e. $\mathcal{A}_1$ or $\mathcal{A}'$) which has the same algebra object $ (1 + \eta)(1 + L_{(1, 2, 2)}) =  (1 + \eta)(1 + L_{(1, 4, 2)})$ is Morita equivalent to $\mathcal{A}_2$ or $\mathcal{A}^*_{E_6}$ depends crucially on the multiplication morphism. The choice of $\mathcal{A}_1$ and the importance of the multiplication morphism was suggested to the authors by Y.~Wang in private communication when we originally proposed the idea that the algebra object should be $\mathcal{A}_2$.} We will implicitly show that it is the case by gauging the $\mathcal{M}(A_{10}, E_{6})$ theory explicitly by using the non-invertible symmetry $\mathcal{A}_2$, resulting in $\mathcal{M}(A_{10}, D_{7})$.


Thus, we will attempt gauging of $\mathcal{M}(A_{10}, E_{6})$ by the algebra $\mathcal{A}_2$. Before the explicit computation, let us offer another motivation to choose it as a suitable algebra. While $\mathcal{M}(A_{10}, D_{7})$ is obtained from the $\mathbb{Z}_2$ gauging of $\mathcal{M}(A_{10}, A_{11})$, the partition function of the former is given by the twisting (\ref{twist}) of the latter $\mathcal{M}(A_{10}, A_{11})$ by an involution. As we will see below, the effect of inserting $L_{(1, 4, 2)}$ along either time or spatial direction on $\mathcal{M}(A_{10}, E_{6})$ is almost the same as inserting $L_{(1,2,2)}$ except for the involution (\ref{twist}).

The next step is to determine the multiplication morphism for the algebra object $\mathcal{A}_2$. An algebra object which takes the form
\begin{align}
    \mathcal{A} = 1 + L_x
\end{align}
is called a binary algebra. Here we assume $L_x$ is self-dual TDL for simplicity. In the case of the multiplicity free $\widetilde{N}_{xx}^{x} = 1$ case, we can derive the constraints on its multiplication morphism from the associativity \cite{Diatlyk:2023fwf}. Let us fix the normalization \cite{Chang:2018iay} by 
\begin{align}
    m_{11}^{1} = m_{1x}^{x} = m_{x1}^{x} = m_{xx}^{1} = \dfrac{1}{\sqrt{\braket{\mathcal{A}}}}.
\end{align}
Then the partition function of the gauged theory is diagrammatically given by 
\small
\begin{align}
    Z_{\mathcal{T}/\mathcal{A}} =&\ \begin{tikzpicture}[scale=0.6,baseline={([yshift=-.5ex]current bounding box.center)},vertex/.style={anchor=base,
    circle,fill=black!25,minimum size=18pt,inner sep=2pt},scale=0.50]        
    \draw(0,0)--(6, 0)--(6,6)--(0,6)--(0,0);
    \draw[->](3,0)--(3.5,1);
    \draw(3.5,1)--(4,2);
    \draw[->](6,3)--(5,2.5);
    \draw(5,2.5)--(4,2);
    \draw[->](4,2)--(3,3);
    \draw(3,3)--(2,4);
    \draw[->](2,4)--(1,3.5);
    \draw(1,3.5)--(0,3);
    \draw[->](2,4)--(2.5,5);
    \draw(2.5,5)--(3,6);
    \draw(2.5,5)node[right]{$\mathcal{A}$};
    \draw(5,2.5)node[above]{$\mathcal{A}$};
    \draw(2.3,2.3)node{$\mathcal{A}$};
    \fill(4,2)circle(0.1);
    \fill(2,4)circle(0.1);
    \draw(2.2,3.8)node[above left]{$m$};
    \draw(3.8,2.2)node[below right]{$m$};
    \end{tikzpicture}\notag\\
    =& \dfrac{1}{\braket{A}}\ \begin{tikzpicture}[scale=0.6,baseline={([yshift=-.5ex]current bounding box.center)},vertex/.style={anchor=base,
    circle,fill=black!25,minimum size=18pt,inner sep=2pt},scale=0.50]        
    \draw(0,0)--(6, 0)--(6,6)--(0,6)--(0,0);
    \draw[dotted,->](3,0)--(3.5,1);
    \draw[dotted](3.5,1)--(4,2);
    \draw[dotted,->](6,3)--(5,2.5);
    \draw[dotted](5,2.5)--(4,2);
    \draw[dotted,->](4,2)--(3,3);
    \draw[dotted](3,3)--(2,4);
    \draw[dotted,->](2,4)--(1,3.5);
    \draw[dotted](1,3.5)--(0,3);
    \draw[dotted,->](2,4)--(2.5,5);
    \draw[dotted](2.5,5)--(3,6);
    \draw(2.5,5)node[right]{$1$};
    \draw(5,2.5)node[above]{$1$};
    \draw(2.3,2.3)node{$1$};
    \end{tikzpicture}
    + \dfrac{1}{\braket{A}}\ \begin{tikzpicture}[scale=0.6,baseline={([yshift=-.5ex]current bounding box.center)},vertex/.style={anchor=base,
    circle,fill=black!25,minimum size=18pt,inner sep=2pt},scale=0.50]        
    \draw(0,0)--(6, 0)--(6,6)--(0,6)--(0,0);
    \draw[->](3,0)--(3.5,1);
    \draw(3.5,1)--(4,2);
    \draw[dotted,->](6,3)--(5,2.5);
    \draw[dotted](5,2.5)--(4,2);
    \draw[->](4,2)--(3,3);
    \draw(3,3)--(2,4);
    \draw[dotted,->](2,4)--(1,3.5);
    \draw[dotted](1,3.5)--(0,3);
    \draw[->](2,4)--(2.5,5);
    \draw(2.5,5)--(3,6);
    \draw(2.5,5)node[right]{$L_x$};
    \draw(5,2.5)node[above]{$1$};
    \draw(2.3,2.3)node{$L_x$};
    \end{tikzpicture}
    + \dfrac{1}{\braket{A}}\ \begin{tikzpicture}[scale=0.6,baseline={([yshift=-.5ex]current bounding box.center)},vertex/.style={anchor=base,
    circle,fill=black!25,minimum size=18pt,inner sep=2pt},scale=0.50]        
    \draw(0,0)--(6, 0)--(6,6)--(0,6)--(0,0);
    \draw[dotted,->](3,0)--(3.5,1);
    \draw[dotted](3.5,1)--(4,2);
    \draw[->](6,3)--(5,2.5);
    \draw(5,2.5)--(4,2);
    \draw[->](4,2)--(3,3);
    \draw(3,3)--(2,4);
    \draw[->](2,4)--(1,3.5);
    \draw(1,3.5)--(0,3);
    \draw[dotted,->](2,4)--(2.5,5);
    \draw[dotted](2.5,5)--(3,6);
    \draw(2.5,5)node[right]{$L_x$};
    \draw(5,2.5)node[above]{$1$};
    \draw(2.3,2.3)node{$L_x$};
    \end{tikzpicture}
    + \dfrac{1}{\braket{A}}\ \begin{tikzpicture}[scale=0.6,baseline={([yshift=-.5ex]current bounding box.center)},vertex/.style={anchor=base,
    circle,fill=black!25,minimum size=18pt,inner sep=2pt},scale=0.50]        
    \draw(0,0)--(6, 0)--(6,6)--(0,6)--(0,0);
    \draw[->](3,0)--(3.5,1);
    \draw(3.5,1)--(4,2);
    \draw[->](6,3)--(5,2.5);
    \draw(5,2.5)--(4,2);
    \draw[dotted,->](4,2)--(3,3);
    \draw[dotted](3,3)--(2,4);
    \draw[->](2,4)--(1,3.5);
    \draw(1,3.5)--(0,3);
    \draw[->](2,4)--(2.5,5);
    \draw(2.5,5)--(3,6);
    \draw(2.5,5)node[right]{$L_x$};
    \draw(5,2.5)node[above]{$L_x$};
    \draw(2.3,2.3)node{$1$};
    \end{tikzpicture}
    + (m_{xx}^{x})^2\ \begin{tikzpicture}[scale=0.6,baseline={([yshift=-.5ex]current bounding box.center)},vertex/.style={anchor=base,
    circle,fill=black!25,minimum size=18pt,inner sep=2pt},scale=0.50]        
    \draw(0,0)--(6, 0)--(6,6)--(0,6)--(0,0);
    \draw[->](3,0)--(3.5,1);
    \draw(3.5,1)--(4,2);
    \draw[->](6,3)--(5,2.5);
    \draw(5,2.5)--(4,2);
    \draw[->](4,2)--(3,3);
    \draw(3,3)--(2,4);
    \draw[->](2,4)--(1,3.5);
    \draw(1,3.5)--(0,3);
    \draw[->](2,4)--(2.5,5);
    \draw(2.5,5)--(3,6);
    \draw(2.5,5)node[right]{$L_x$};
    \draw(5,2.5)node[above]{$L_x$};
    \draw(2.3,2.3)node{$L_x$};
    \end{tikzpicture}\notag\\
    = &\dfrac{1}{\braket{A}}\ \begin{tikzpicture}[scale=0.6,baseline={([yshift=-.5ex]current bounding box.center)},vertex/.style={anchor=base,
    circle,fill=black!25,minimum size=18pt,inner sep=2pt},scale=0.50]        
    \draw(0,0)--(6, 0)--(6,6)--(0,6)--(0,0);
    \draw[dotted,->](3,0)--(2.5,1);
    \draw[dotted](2.5,1)--(2,2);
    \draw[dotted](0,3)--(1,2.5);
    \draw[dotted,->](2,2)--(1,2.5);
    \draw[dotted,->](2,2)--(3,3);
    \draw[dotted](3,3)--(4,4);
    \draw[dotted,->](4,4)--(3.5,5);
    \draw[dotted](3.5,5)--(3,6);
    \draw[dotted,->](6,3)--(5,3.5);
    \draw[dotted](5,3.5)--(4,4);
    \draw(1,2.3)node[above]{$1$};
    \draw(3.8,5)node[left]{$1$};
    \draw(3.7,2.3)node{$1$};
    \end{tikzpicture}
    + \dfrac{1}{\braket{A}}\ \begin{tikzpicture}[scale=0.6,baseline={([yshift=-.5ex]current bounding box.center)},vertex/.style={anchor=base,
    circle,fill=black!25,minimum size=18pt,inner sep=2pt},scale=0.50]        
    \draw(0,0)--(6, 0)--(6,6)--(0,6)--(0,0);
    \draw[->](3,0)--(2.5,1);
    \draw(2.5,1)--(2,2);
    \draw[dotted](0,3)--(1,2.5);
    \draw[dotted,->](2,2)--(1,2.5);
    \draw[->](2,2)--(3,3);
    \draw(3,3)--(4,4);
    \draw[->](4,4)--(3.5,5);
    \draw(3.5,5)--(3,6);
    \draw[dotted,->](6,3)--(5,3.5);
    \draw[dotted](5,3.5)--(4,4);
    \draw(1,2.3)node[above]{$1$};
    \draw(3.8,5)node[left]{$L_x$};
    \draw(3.7,2.3)node{$L_x$};
    \end{tikzpicture}
    + \dfrac{1}{\braket{A}}\ \begin{tikzpicture}[scale=0.6,baseline={([yshift=-.5ex]current bounding box.center)},vertex/.style={anchor=base,
    circle,fill=black!25,minimum size=18pt,inner sep=2pt},scale=0.50]        
    \draw(0,0)--(6, 0)--(6,6)--(0,6)--(0,0);
    \draw[dotted,->](3,0)--(2.5,1);
    \draw[dotted](2.5,1)--(2,2);
    \draw(0,3)--(1,2.5);
    \draw[->](2,2)--(1,2.5);
    \draw[->](2,2)--(3,3);
    \draw(3,3)--(4,4);
    \draw[dotted,->](4,4)--(3.5,5);
    \draw[dotted](3.5,5)--(3,6);
    \draw[->](6,3)--(5,3.5);
    \draw(5,3.5)--(4,4);
    \draw(1,2.3)node[above]{$L_x$};
    \draw(3.8,5)node[left]{$1$};
    \draw(3.7,2.3)node{$L_x$};
    \end{tikzpicture}\notag\\
    &+ \left(\dfrac{F_{xxx}^{x}(1, 1)}{\braket{A}} + F_{xxx}^{x}(L_x, 1)(m_{xx}^{x})^2\right)\ \begin{tikzpicture}[scale=0.6,baseline={([yshift=-.5ex]current bounding box.center)},vertex/.style={anchor=base,
    circle,fill=black!25,minimum size=18pt,inner sep=2pt},scale=0.50]        
    \draw(0,0)--(6, 0)--(6,6)--(0,6)--(0,0);
    \draw[->](3,0)--(2.5,1);
    \draw(2.5,1)--(2,2);
    \draw[dotted](0,3)--(1,2.5);
    \draw[dotted,->](2,2)--(1,2.5);
    \draw[->](2,2)--(3,3);
    \draw(3,3)--(4,4);
    \draw[->](4,4)--(3.5,5);
    \draw(3.5,5)--(3,6);
    \draw[dotted,->](6,3)--(5,3.5);
    \draw[dotted](5,3.5)--(4,4);
    \draw(1,2.3)node[above]{$1$};
    \draw(3.8,5)node[left]{$L_x$};
    \draw(3.7,2.3)node{$L_x$};
    \end{tikzpicture}
    + \left(\dfrac{F_{xxx}^{x}(1, L_x)}{\braket{A}} + F_{xxx}^{x}(L_x, L_x)(m_{xx}^{x})^2\right)\ \begin{tikzpicture}[scale=0.6,baseline={([yshift=-.5ex]current bounding box.center)},vertex/.style={anchor=base,
    circle,fill=black!25,minimum size=18pt,inner sep=2pt},scale=0.50]        
    \draw(0,0)--(6, 0)--(6,6)--(0,6)--(0,0);
    \draw[->](3,0)--(2.5,1);
    \draw(2.5,1)--(2,2);
    \draw(0,3)--(1,2.5);
    \draw[->](2,2)--(1,2.5);
    \draw[->](2,2)--(3,3);
    \draw(3,3)--(4,4);
    \draw[->](4,4)--(3.5,5);
    \draw(3.5,5)--(3,6);
    \draw[->](6,3)--(5,3.5);
    \draw(5,3.5)--(4,4);
    \draw(1,2.3)node[above]{$L_x$};
    \draw(3.8,5)node[left]{$L_x$};
    \draw(3.7,2.3)node{$L_x$};
    \end{tikzpicture}\notag\\
    &+ \sum_{L_y \neq 1, L_x} \left(\dfrac{F_{xxx}^{x}(1, L_y)}{\braket{A}} + F_{xxx}^{x}(L_x, L_y)(m_{xx}^{x})^2\right)\ \begin{tikzpicture}[scale=0.6,baseline={([yshift=-.5ex]current bounding box.center)},vertex/.style={anchor=base,
    circle,fill=black!25,minimum size=18pt,inner sep=2pt},scale=0.50]        
    \draw(0,0)--(6, 0)--(6,6)--(0,6)--(0,0);
    \draw[->](3,0)--(2.5,1);
    \draw(2.5,1)--(2,2);
    \draw(0,3)--(1,2.5);
    \draw[->](2,2)--(1,2.5);
    \draw[->](2,2)--(3,3);
    \draw(3,3)--(4,4);
    \draw[->](4,4)--(3.5,5);
    \draw(3.5,5)--(3,6);
    \draw[->](6,3)--(5,3.5);
    \draw(5,3.5)--(4,4);
    \draw(1,2.3)node[above]{$L_x$};
    \draw(3.8,5)node[left]{$L_x$};
    \draw(3.7,2.3)node{$L_y$};
    \end{tikzpicture}\label{T/A1}
\end{align}
\normalsize
On the other hand, from the associativity, we can also rewrite the partition function of the gauged theory as
\begin{align}
    Z_{\mathcal{T}/\mathcal{A}} =\ &\begin{tikzpicture}[scale=0.6,baseline={([yshift=-.5ex]current bounding box.center)},vertex/.style={anchor=base,
    circle,fill=black!25,minimum size=18pt,inner sep=2pt},scale=0.50]        
    \draw(0,0)--(6, 0)--(6,6)--(0,6)--(0,0);
    \draw[->](3,0)--(2.5,1);
    \draw(2.5,1)--(2,2);
    \draw(0,3)--(1,2.5);
    \draw[->](2,2)--(1,2.5);
    \draw[->](2,2)--(3,3);
    \draw(3,3)--(4,4);
    \draw[->](4,4)--(3.5,5);
    \draw(3.5,5)--(3,6);
    \draw[->](6,3)--(5,3.5);
    \draw(5,3.5)--(4,4);
    \draw(1,2.3)node[above]{$\mathcal{A}$};
    \draw(3.8,5)node[left]{$\mathcal{A}$};
    \draw(3.7,2.3)node{$\mathcal{A}$};
    \fill(2,2)circle(0.1);
    \fill(4,4)circle(0.1);
    \draw(2.2,2.2)node[below left]{$m$};
    \draw(3.8,3.8)node[above right]{$m$};
    \end{tikzpicture}\notag\\
    = &\dfrac{1}{\braket{A}}\ \begin{tikzpicture}[scale=0.6,baseline={([yshift=-.5ex]current bounding box.center)},vertex/.style={anchor=base,
    circle,fill=black!25,minimum size=18pt,inner sep=2pt},scale=0.50]        
    \draw(0,0)--(6, 0)--(6,6)--(0,6)--(0,0);
    \draw[dotted,->](3,0)--(2.5,1);
    \draw[dotted](2.5,1)--(2,2);
    \draw[dotted](0,3)--(1,2.5);
    \draw[dotted,->](2,2)--(1,2.5);
    \draw[dotted,->](2,2)--(3,3);
    \draw[dotted](3,3)--(4,4);
    \draw[dotted,->](4,4)--(3.5,5);
    \draw[dotted](3.5,5)--(3,6);
    \draw[dotted,->](6,3)--(5,3.5);
    \draw[dotted](5,3.5)--(4,4);
    \draw(1,2.3)node[above]{$1$};
    \draw(3.8,5)node[left]{$1$};
    \draw(3.7,2.3)node{$1$};
    \end{tikzpicture}
    + \dfrac{1}{\braket{A}}\ \begin{tikzpicture}[scale=0.6,baseline={([yshift=-.5ex]current bounding box.center)},vertex/.style={anchor=base,
    circle,fill=black!25,minimum size=18pt,inner sep=2pt},scale=0.50]        
    \draw(0,0)--(6, 0)--(6,6)--(0,6)--(0,0);
    \draw[->](3,0)--(2.5,1);
    \draw(2.5,1)--(2,2);
    \draw[dotted](0,3)--(1,2.5);
    \draw[dotted,->](2,2)--(1,2.5);
    \draw[->](2,2)--(3,3);
    \draw(3,3)--(4,4);
    \draw[->](4,4)--(3.5,5);
    \draw(3.5,5)--(3,6);
    \draw[dotted,->](6,3)--(5,3.5);
    \draw[dotted](5,3.5)--(4,4);
    \draw(1,2.3)node[above]{$1$};
    \draw(3.8,5)node[left]{$L_x$};
    \draw(3.7,2.3)node{$L_x$};
    \end{tikzpicture}
    + \dfrac{1}{\braket{A}}\ \begin{tikzpicture}[scale=0.6,baseline={([yshift=-.5ex]current bounding box.center)},vertex/.style={anchor=base,
    circle,fill=black!25,minimum size=18pt,inner sep=2pt},scale=0.50]        
    \draw(0,0)--(6, 0)--(6,6)--(0,6)--(0,0);
    \draw[dotted,->](3,0)--(2.5,1);
    \draw[dotted](2.5,1)--(2,2);
    \draw(0,3)--(1,2.5);
    \draw[->](2,2)--(1,2.5);
    \draw[->](2,2)--(3,3);
    \draw(3,3)--(4,4);
    \draw[dotted,->](4,4)--(3.5,5);
    \draw[dotted](3.5,5)--(3,6);
    \draw[->](6,3)--(5,3.5);
    \draw(5,3.5)--(4,4);
    \draw(1,2.3)node[above]{$L_x$};
    \draw(3.8,5)node[left]{$1$};
    \draw(3.7,2.3)node{$L_x$};
    \end{tikzpicture}
    + \dfrac{1}{\braket{A}}\ \begin{tikzpicture}[scale=0.6,baseline={([yshift=-.5ex]current bounding box.center)},vertex/.style={anchor=base,
    circle,fill=black!25,minimum size=18pt,inner sep=2pt},scale=0.50]        
    \draw(0,0)--(6, 0)--(6,6)--(0,6)--(0,0);
    \draw[->](3,0)--(2.5,1);
    \draw(2.5,1)--(2,2);
    \draw(0,3)--(1,2.5);
    \draw[->](2,2)--(1,2.5);
    \draw[dotted,->](2,2)--(3,3);
    \draw[dotted](3,3)--(4,4);
    \draw[->](4,4)--(3.5,5);
    \draw(3.5,5)--(3,6);
    \draw[->](6,3)--(5,3.5);
    \draw(5,3.5)--(4,4);
    \draw(1,2.3)node[above]{$L_x$};
    \draw(3.8,5)node[left]{$1$};
    \draw(3.7,2.3)node{$L_x$};
    \end{tikzpicture}    
    + (m_{xx}^{x})^2\ \begin{tikzpicture}[scale=0.6,baseline={([yshift=-.5ex]current bounding box.center)},vertex/.style={anchor=base,
    circle,fill=black!25,minimum size=18pt,inner sep=2pt},scale=0.50]        
    \draw(0,0)--(6, 0)--(6,6)--(0,6)--(0,0);
    \draw[->](3,0)--(2.5,1);
    \draw(2.5,1)--(2,2);
    \draw(0,3)--(1,2.5);
    \draw[->](2,2)--(1,2.5);
    \draw[->](2,2)--(3,3);
    \draw(3,3)--(4,4);
    \draw[->](4,4)--(3.5,5);
    \draw(3.5,5)--(3,6);
    \draw[->](6,3)--(5,3.5);
    \draw(5,3.5)--(4,4);
    \draw(1,2.3)node[above]{$L_x$};
    \draw(3.8,5)node[left]{$L_x$};
    \draw(3.7,2.3)node{$L_x$};
    \end{tikzpicture}\label{T/A2}
\end{align}
Comparing (\ref{T/A1}) and (\ref{T/A2}), we get the relation between the multiplication morphism and F-symbols
\begin{equation}
    \begin{split}
        F_{xxx}^{x}(1, 1) + \braket{\mathcal{A}}F_{xxx}^{x}(L_x, 1)(m_{xx}^{x})^2 = 1\\
        F_{xxx}^{x}(1, L_x) + \braket{\mathcal{A}}F_{xxx}^{x}(L_x, L_x)(m_{xx}^{x})^2 = \braket{A}(m_{xx}^{x})^2\\
        F_{xxx}^{x}(1, L_y) + \braket{\mathcal{A}}F_{xxx}^{x}(L_x, L_y)(m_{xx}^{x})^2 = 0 , \label{mmF}
    \end{split}
\end{equation}

where $L_{y} \neq 1, L_x$. To proceed, we work in the conventional gauge
\begin{align}
    F_{L_{(1, 4, 2)}L_{(1, 4, 2)}L_{(1, 4, 2)}}^{L_{(1, 4, 2)}}(L_{(1, 4, 2)}, 1) = 1,\  F_{L_{(1, 4, 2)}L_{(1, 4, 2)}L_{(1, 4, 2)}}^{L_{(1, 4, 2)}}(1, 1) = \dfrac{1}{\braket{L_{(1, 4, 2)}}} = 2 - \sqrt{3}.
\end{align}
Then the multiplication morphism for the binary algebra $\mathcal{A}_2$ is
\begin{align}
    m_{11}^{1} = m_{1\ L_{(1, 4, 2)}}^{L_{(1, 4, 2)}} = m_{L_{(1, 4, 2)}\ 1}^{L_{(1, 4, 2)}} = m_{L_{(1, 4, 2)}\ L_{(1, 4, 2)}}^{1} = \dfrac{1}{\braket{\mathcal{A}_2}} = \dfrac{1}{\sqrt{3 + \sqrt{3}}},\ m_{L_{(1, 4, 2)}\ L_{(1, 4, 2)}}^{L_{(1, 4, 2)}} = \sqrt{\dfrac{\sqrt{3} - 1}{3 + \sqrt{3}}}\label{m}
\end{align}
from the first line of (\ref{mmF}). We note that they are the same as those for $\mathcal{A}_{E_6}^*$ obtained in \cite{Diatlyk:2023fwf}.

Now we are in the position to compute the partition function of the gauged theory. From here, we denote $1, L_{(1, 2, 2)}$ and $L_{(1, 4, 2)}$ as the dotted lines, the red lines and the blue lines, respectively. We start with $\mathcal{A}_{E_6}^*$ and then $\mathcal{A}_2$ in comparison.

The untwisted partition function of $\mathcal{M}(A_{10}, E_6)$ is 
\begin{align}
    \begin{tikzpicture}[scale=0.6,baseline={([yshift=-.5ex]current bounding box.center)},vertex/.style={anchor=base,
    circle,fill=black!25,minimum size=18pt,inner sep=2pt},scale=0.50] 
    \draw(0,0)--(6, 0)--(6,6)--(0,6)--(0,0);
    \draw[dotted,->](3,0)--(3.5,1);
    \draw[dotted](3.5,1)--(4,2);
    \draw[dotted,->](6,3)--(5,2.5);
    \draw[dotted](5,2.5)--(4,2);
    \draw[dotted,->](4,2)--(3,3);
    \draw[dotted](3,3)--(2,4);
    \draw[dotted,->](2,4)--(1,3.5);
    \draw[dotted](1,3.5)--(0,3);
    \draw[dotted,->](2,4)--(2.5,5);
    \draw[dotted](2.5,5)--(3,6);
    \end{tikzpicture} = \sum_{\substack{r = 1\\ \mathrm{step}\ 2}}^{9}\left\{\lvert\chi_{r, 1} + \chi_{r, 7}\rvert^2 + \lvert\chi_{r, 4} + \chi_{r, 8}\rvert^2 + \lvert\chi_{r, 5} + \chi_{r, 11}\rvert^2\right\}.\label{A10E6pf}
\end{align}
From the result by Petkova and Zuber (\ref{twistedpartition}), the partition function twisted by $L_{(1, 2, 2)}$ is 
\begin{align}
    \begin{tikzpicture}[scale=0.6,baseline={([yshift=-.5ex]current bounding box.center)},vertex/.style={anchor=base,
    circle,fill=black!25,minimum size=18pt,inner sep=2pt},scale=0.50]        
    \draw(0,0)--(6, 0)--(6,6)--(0,6)--(0,0);
    \draw[red, ->](3,0)--(3.5,1);
    \draw[red](3.5,1)--(4,2);
    \draw[dotted,->](6,3)--(5,2.5);
    \draw[dotted](5,2.5)--(4,2);
    \draw[red, ->](4,2)--(3,3);
    \draw[red](3,3)--(2,4);
    \draw[dotted,->](2,4)--(1,3.5);
    \draw[dotted](1,3.5)--(0,3);
    \draw[red, ->](2,4)--(2.5,5);
    \draw[red](2.5,5)--(3,6);
    \end{tikzpicture} = \sum_{\substack{r = 1\\ \mathrm{step}\ 2}}^{9} \left\{\lvert\chi_{r, 2} + \chi_{r, 6} + \chi_{r, 8}\lvert^2 + \lvert\chi_{r, 4} + \chi_{r, 6} + \chi_{r, 10}\rvert^2 + \lvert\chi_{r, 3} + \chi_{r, 5} + \chi_{r, 7} + \chi_{r, 9}\rvert^2\right\}.\label{122zu1}
\end{align}
We can now perform the modular S-transformation on (\ref{122zu1}) to obtain
\begin{align}
    \begin{tikzpicture}[scale=0.6,baseline={([yshift=-.5ex]current bounding box.center)},vertex/.style={anchor=base,
    circle,fill=black!25,minimum size=18pt,inner sep=2pt},scale=0.50]        
    \draw(0,0)--(6, 0)--(6,6)--(0,6)--(0,0);
    \draw[dotted,->](3,0)--(3.5,1);
    \draw[dotted](3.5,1)--(4,2);
    \draw[red, ->](6,3)--(5,2.5);
    \draw[red](5,2.5)--(4,2);
    \draw[red, ->](4,2)--(3,3);
    \draw[red](3,3)--(2,4);
    \draw[red, ->](2,4)--(1,3.5);
    \draw[red](1,3.5)--(0,3);
    \draw[dotted,->](2,4)--(2.5,5);
    \draw[dotted](2.5,5)--(3,6);
    \end{tikzpicture} = \sum_{\substack{r = 1\\ \mathrm{step}\ 2}}^{9} \left\{\left\lvert\dfrac{1 + \sqrt{3}}{\sqrt{2}}\chi_{r, 1} + \dfrac{1 - \sqrt{3}}{\sqrt{2}}\chi_{r, 7}\right\rvert^2 + \lvert\chi_{r, 4} - \chi_{r, 8}\rvert^2 + \left\lvert\dfrac{1 - \sqrt{3}}{\sqrt{2}}\chi_{r, 5} + \dfrac{1 + \sqrt{3}}{\sqrt{2}}\chi_{r, 11}\right\rvert^2\right\}.\label{122zu2}
\end{align}
On the other hand, we can perform the modular $T^{-1}$-transformation to obtain
\begin{align}
    \begin{tikzpicture}[scale=0.6,baseline={([yshift=-.5ex]current bounding box.center)},vertex/.style={anchor=base,
    circle,fill=black!25,minimum size=18pt,inner sep=2pt},scale=0.50]        
    \draw(0,0)--(6, 0)--(6,6)--(0,6)--(0,0);
    \draw[red, ->](3,0)--(3.5,1);
    \draw[red](3.5,1)--(4,2);
    \draw[red, ->](6,3)--(5,2.5);
    \draw[red](5,2.5)--(4,2);
    \draw[dotted,->](4,2)--(3,3);
    \draw[dotted](3,3)--(2,4);
    \draw[red, ->](2,4)--(1,3.5);
    \draw[red](1,3.5)--(0,3);
    \draw[red, ->](2,4)--(2.5,5);
    \draw[red](2.5,5)--(3,6);
    \end{tikzpicture} = \sum_{\substack{r = 1\\ \mathrm{step}\ 2}}^{9} (&\chi_{r, 2}\bar{\chi}_{r, 2} + \omega\chi_{r, 2}\bar{\chi}_{r, 6} - i\chi_{r, 2}\bar{\chi}_{r, 8} + \chi_{r, 3}\bar{\chi}_{r, 3} + \omega^2\chi_{r, 3}\bar{\chi}_{r, 5} - \omega^2\chi_{r, 3}\bar{\chi}_{r, 7} - \chi_{r, 3}\bar{\chi}_{r, 9}\notag\\
    &+ \chi_{r, 4}\bar{\chi}_{r, 4} + e^{\frac{7}{6}\pi i}\chi_{r, 4}\bar{\chi}_{r, 6} + i\chi_{r, 4}\bar{\chi}_{r, 10} + \omega\chi_{r, 5}\bar{\chi}_{r, 3} + \chi_{r, 5}\bar{\chi}_{r, 5} - \chi_{r, 5}\bar{\chi}_{r, 7} - \omega\chi_{r, 5}\bar{\chi}_{r, 9}\notag\\
    &+ \omega^2\chi_{r, 6}\bar{\chi}_{r, 2} + e^{\frac{5}{6}\pi i}\chi_{r, 6}\bar{\chi}_{r, 4} + 2\chi_{r, 6}\bar{\chi}_{r, 6} +  e^{\frac{5}{6}\pi i}\chi_{r, 6}\bar{\chi}_{r, 8} + \omega^2\chi_{r, 6}\bar{\chi}_{r, 10}\notag\\
    &- \omega\chi_{r, 7}\bar{\chi}_{r, 3} - \chi_{r, 7}\bar{\chi}_{r, 5} + \chi_{r, 7}\bar{\chi}_{r, 7} + \omega\chi_{r, 7}\bar{\chi}_{r, 9} + i\chi_{r, 8}\bar{\chi}_{r, 2} + e^{\frac{7}{6}\pi i}\chi_{r, 8}\bar{\chi}_{r, 6} + \chi_{r, 8}\bar{\chi}_{r, 8}\notag\\
    &- \chi_{r, 9}\bar{\chi}_{r, 3} - \omega^2\chi_{r, 9}\bar{\chi}_{r, 5} + \omega^2\chi_{r, 9}\bar{\chi}_{r, 7} + \chi_{r, 9}\bar{\chi}_{r, 9} - i\chi_{r, 10}\bar{\chi}_{r, 4} + \omega\chi_{r, 10}\bar{\chi}_{r, 6} + \chi_{r, 10}\bar{\chi}_{r, 10}),\label{122zu3}
\end{align}
where $\omega = e^{\frac{2}{3}\pi i}$.
The last diagram is the most complicated. The results in \cite{Diatlyk:2023fwf} show
\begin{align}
    \begin{tikzpicture}[scale=0.6,baseline={([yshift=-.5ex]current bounding box.center)},vertex/.style={anchor=base,
    circle,fill=black!25,minimum size=18pt,inner sep=2pt},scale=0.50]        
    \draw(0,0)--(6, 0)--(6,6)--(0,6)--(0,0);
    \draw[red, ->](3,0)--(3.5,1);
    \draw[red](3.5,1)--(4,2);
    \draw[red, ->](6,3)--(5,2.5);
    \draw[red](5,2.5)--(4,2);
    \draw[red, ->](4,2)--(3,3);
    \draw[red](3,3)--(2,4);
    \draw[red, ->](2,4)--(1,3.5);
    \draw[red](1,3.5)--(0,3);
    \draw[red, ->](2,4)--(2.5,5);
    \draw[red](2.5,5)--(3,6);
    \end{tikzpicture} = &\dfrac{1}{\sqrt{3} - 1}\sum_{\substack{r = 1\\ \mathrm{step}\ 2}}^{9} \{(1 + \sqrt{3})\chi_{r, 2}\bar{\chi}_{r, 2} + \omega^2\chi_{r, 2}\bar{\chi}_{r, 6} + (-1 + i)\chi_{r, 2}\bar{\chi}_{r, 8}\notag\\
    &+ (1 + \sqrt{3})\chi_{r, 3}\bar{\chi}_{r, 3} + \omega\chi_{r, 3}\bar{\chi}_{r, 5} + (-1 + \omega^2)\chi_{r, 3}\bar{\chi}_{r, 7}\notag\\
    &+ (-1 + \sqrt{3})\chi_{r, 4}\bar{\chi}_{r, 4} + (-1 - e^{\frac{7}{6}\pi i})\chi_{r, 4}\bar{\chi}_{r, 6} + (-1 - i)\chi_{r, 4}\bar{\chi}_{r, 10}\notag\\
    &+ \omega^2\chi_{r, 5}\bar{\chi}_{r, 3} + (-2 + 2\sqrt{3})\chi_{r, 5}\bar{\chi}_{r, 5} + (-1 + \omega)\chi_{r, 5}\bar{\chi}_{r, 9}\notag\\
    &+ \omega\chi_{r, 6}\bar{\chi}_{r, 2} + (-1 - e^{\frac{5}{6}\pi i})\chi_{r, 6}\bar{\chi}_{r, 4} + (-1 + \sqrt{3})\chi_{r, 6}\bar{\chi}_{r, 6} + (-1 - e^{\frac{5}{6}\pi i})\chi_{r, 6}\bar{\chi}_{r, 8} + \omega\chi_{r, 6}\bar{\chi}_{r, 10}\notag\\
    &+ (-1 + \omega)\chi_{r, 7}\bar{\chi}_{r, 3} + (-2 + 2\sqrt{3})\chi_{r, 7}\bar{\chi}_{r, 7} + \omega^2\chi_{r, 7}\bar{\chi}_{r, 9}\notag\\
    &+ (-1 - i)\chi_{r, 8}\bar{\chi}_{r, 2} + (-1 - e^{\frac{7}{6}\pi i})\chi_{r, 8}\bar{\chi}_{r, 6} + (-1 + \sqrt{3})\chi_{r, 8}\bar{\chi}_{r, 8}\notag\\
    &+ (-1 + \omega^2)\chi_{r, 9}\bar{\chi}_{r, 5} + \omega\chi_{r, 9}\bar{\chi}_{r, 7} + (1 + \sqrt{3})\chi_{r, 9}\bar{\chi}_{r, 9}\notag\\
    &+ (-1 + i)\chi_{r, 10}\bar{\chi}_{r, 4} + \omega^2\chi_{r, 10}\bar{\chi}_{r, 6} + (1 + \sqrt{3})\chi_{r, 10}\bar{\chi}_{r, 10}\}.\label{122zu4}
\end{align}
We see that the sum of all the contributions with weight specified by the multiplication morphisms gives the partition function of $\mathcal{M}(A_{10},A_{11})$.

Let us now proceed to our main interest of gauging $\mathcal{A}_2$.
The untwisted partition function of $\mathcal{M}(A_{10}, E_6)$ is the same as (\ref{A10E6pf}).

From the result by Petkova and Zuber (\ref{twistedpartition}), the partition function twisted by $L_{(1, 4, 2)}$ is 
\begin{align}
    \begin{tikzpicture}[scale=0.6,baseline={([yshift=-.5ex]current bounding box.center)},vertex/.style={anchor=base,
    circle,fill=black!25,minimum size=18pt,inner sep=2pt},scale=0.50]        
    \draw(0,0)--(6, 0)--(6,6)--(0,6)--(0,0);
    \draw[blue, ->](3,0)--(3.5,1);
    \draw[blue](3.5,1)--(4,2);
    \draw[dotted,->](6,3)--(5,2.5);
    \draw[dotted](5,2.5)--(4,2);
    \draw[blue, ->](4,2)--(3,3);
    \draw[blue](3,3)--(2,4);
    \draw[dotted,->](2,4)--(1,3.5);
    \draw[dotted](1,3.5)--(0,3);
    \draw[blue, ->](2,4)--(2.5,5);
    \draw[blue](2.5,5)--(3,6);
    \end{tikzpicture} = &\sum_{\substack{r = 1\\ \mathrm{step}\ 2}}^{9} \{(\chi_{r, 2} + \chi_{r, 6} + \chi_{r, 8})(\bar{\chi}_{r, 4} + \bar{\chi}_{r, 6} + \bar{\chi}_{r, 10}) + (\chi_{r, 4} + \chi_{r, 6} + \chi_{r, 10})(\bar{\chi}_{r, 2} + \bar{\chi}_{r, 6} + \bar{\chi}_{r, 8})\notag\\
    &+ \lvert\chi_{r, 3} + \chi_{r, 5} + \chi_{r, 7} + \chi_{r, 9}\rvert^2\}.\label{142zu1}
\end{align}
We can perform the modular S-transformation on (\ref{142zu1}) to obtain
\begin{align}
    \begin{tikzpicture}[scale=0.6,baseline={([yshift=-.5ex]current bounding box.center)},vertex/.style={anchor=base,
    circle,fill=black!25,minimum size=18pt,inner sep=2pt},scale=0.50]        
    \draw(0,0)--(6, 0)--(6,6)--(0,6)--(0,0);
    \draw[dotted,->](3,0)--(3.5,1);
    \draw[dotted](3.5,1)--(4,2);
    \draw[blue, ->](6,3)--(5,2.5);
    \draw[blue](5,2.5)--(4,2);
    \draw[blue, ->](4,2)--(3,3);
    \draw[blue](3,3)--(2,4);
    \draw[blue, ->](2,4)--(1,3.5);
    \draw[blue](1,3.5)--(0,3);
    \draw[dotted,->](2,4)--(2.5,5);
    \draw[dotted](2.5,5)--(3,6);
    \end{tikzpicture} = \sum_{\substack{r = 1\\ \mathrm{step}\ 2}}^{9} \left\{\left\lvert\dfrac{1 + \sqrt{3}}{\sqrt{2}}\chi_{r, 1} + \dfrac{1 - \sqrt{3}}{\sqrt{2}}\chi_{r, 7}\right\rvert^2 - \lvert\chi_{r, 4} - \chi_{r, 8}\rvert^2 + \left\lvert\dfrac{1 - \sqrt{3}}{\sqrt{2}}\chi_{r, 5} + \dfrac{1 + \sqrt{3}}{\sqrt{2}}\chi_{r, 11}\right\rvert^2\right\}.\label{142zu2}
\end{align}
On the other hand, we can perform the modular $T^{-1}$-transformation to obtain
\begin{align}
    \begin{tikzpicture}[scale=0.6,baseline={([yshift=-.5ex]current bounding box.center)},vertex/.style={anchor=base,
    circle,fill=black!25,minimum size=18pt,inner sep=2pt},scale=0.50]        
    \draw(0,0)--(6, 0)--(6,6)--(0,6)--(0,0);
    \draw[blue, ->](3,0)--(3.5,1);
    \draw[blue](3.5,1)--(4,2);
    \draw[blue, ->](6,3)--(5,2.5);
    \draw[blue](5,2.5)--(4,2);
    \draw[dotted,->](4,2)--(3,3);
    \draw[dotted](3,3)--(2,4);
    \draw[blue, ->](2,4)--(1,3.5);
    \draw[blue](1,3.5)--(0,3);
    \draw[blue, ->](2,4)--(2.5,5);
    \draw[blue](2.5,5)--(3,6);
    \end{tikzpicture} = \sum_{\substack{r = 1\\ \mathrm{step}\ 2}}^{9} (&-i\chi_{r, 2}\bar{\chi}_{r, 4} + \omega\chi_{r, 2}\bar{\chi}_{r, 6} + \chi_{r, 2}\bar{\chi}_{r, 10} + \chi_{r, 3}\bar{\chi}_{r, 3} + \omega^2\chi_{r, 3}\bar{\chi}_{r, 5} - \omega^2\chi_{r, 3}\bar{\chi}_{r, 7} - \chi_{r, 3}\bar{\chi}_{r, 9}\notag\\
    &+ i\chi_{r, 4}\bar{\chi}_{r, 2} + e^{\frac{7}{6}\pi i}\chi_{r, 4}\bar{\chi}_{r, 6} + \chi_{r, 4}\bar{\chi}_{r, 8} + \omega\chi_{r, 5}\bar{\chi}_{r, 3} + \chi_{r, 5}\bar{\chi}_{r, 5} - \chi_{r, 5}\bar{\chi}_{r, 7} - \omega\chi_{r, 5}\bar{\chi}_{r, 9}\notag\\
    &+ \omega^2\chi_{r, 6}\bar{\chi}_{r, 2} + e^{\frac{5}{6}\pi i}\chi_{r, 6}\bar{\chi}_{r, 4} + 2\chi_{r, 6}\bar{\chi}_{r, 6} +  e^{\frac{5}{6}\pi i}\chi_{r, 6}\bar{\chi}_{r, 8} + \omega^2\chi_{r, 6}\bar{\chi}_{r, 10}\notag\\
    &+ \omega^2\chi_{r, 7}\bar{\chi}_{r, 3} - \chi_{r, 7}\bar{\chi}_{r, 5} + \chi_{r, 7}\bar{\chi}_{r, 7} + \omega\chi_{r, 7}\bar{\chi}_{r, 9} + \chi_{r, 8}\bar{\chi}_{r, 4} + e^{\frac{7}{6}\pi i}\chi_{r, 8}\bar{\chi}_{r, 6} + i\chi_{r, 8}\bar{\chi}_{r, 10}\notag\\
    &- \chi_{r, 9}\bar{\chi}_{r, 3} - \omega^2\chi_{r, 9}\bar{\chi}_{r, 5} + \omega^2\chi_{r, 9}\bar{\chi}_{r, 7} + \chi_{r, 9}\bar{\chi}_{r, 9} + \chi_{r, 10}\bar{\chi}_{r, 2} + \omega\chi_{r, 10}\bar{\chi}_{r, 6} - i\chi_{r, 10}\bar{\chi}_{r, 8}).\label{142zu3}
\end{align}

Let us make an important observation here. As alluded above, (\ref{142zu1})(\ref{142zu2})(\ref{142zu3}) can be obtained by the involution
\begin{align}
    s \mapsto \zeta(s) = \begin{cases}
        s\ \ \ \ \ \ \ \ \ \ \ \mathrm{for\ odd}\ s\\
        12 - s\ \ \ \ \mathrm{for\ even}\ s
    \end{cases} \label{eventwist}
\end{align}
acting on the Virasoro characters from (\ref{122zu1})(\ref{122zu2})(\ref{122zu3}) respectively. Namely, the twisted partition functions with the TDL $L_{(1,4,2)}$ can be obtained from the twisted partition functions with the $L_{(1,2,2)}$ by using \eqref{eventwist}.

Thus, we apply this involution to (\ref{122zu3}) to obtain 
\begin{align}
    \begin{tikzpicture}[scale=0.6,baseline={([yshift=-.5ex]current bounding box.center)},vertex/.style={anchor=base,
    circle,fill=black!25,minimum size=18pt,inner sep=2pt},scale=0.50]        
    \draw(0,0)--(6, 0)--(6,6)--(0,6)--(0,0);
    \draw[blue, ->](3,0)--(3.5,1);
    \draw[blue](3.5,1)--(4,2);
    \draw[blue, ->](6,3)--(5,2.5);
    \draw[blue](5,2.5)--(4,2);
    \draw[blue, ->](4,2)--(3,3);
    \draw[blue](3,3)--(2,4);
    \draw[blue, ->](2,4)--(1,3.5);
    \draw[blue](1,3.5)--(0,3);
    \draw[blue, ->](2,4)--(2.5,5);
    \draw[blue](2.5,5)--(3,6);
    \end{tikzpicture} = &\dfrac{1}{\sqrt{3} - 1}\sum_{\substack{r = 1\\ \mathrm{step}\ 2}}^{9} \{(-1 + i)\chi_{r, 2}\bar{\chi}_{r, 4} + \omega^2\chi_{r, 2}\bar{\chi}_{r, 6} + (1 + \sqrt{3})\chi_{r, 2}\bar{\chi}_{r, 10}\notag\\
    &+ (1 + \sqrt{3})\chi_{r, 3}\bar{\chi}_{r, 3} + \omega\chi_{r, 3}\bar{\chi}_{r, 5} + (-1 + \omega^2)\chi_{r, 3}\bar{\chi}_{r, 7}\notag\\
    &+ (-1 - i)\chi_{r, 4}\bar{\chi}_{r, 2} + (-1 - e^{\frac{7}{6}\pi i})\chi_{r, 4}\bar{\chi}_{r, 6} + (-1 + \sqrt{3})\chi_{r, 4}\bar{\chi}_{r, 8}\notag\\
    &+ \omega^2\chi_{r, 5}\bar{\chi}_{r, 3} + (-2 + 2\sqrt{3})\chi_{r, 5}\bar{\chi}_{r, 5} + (-1 + \omega)\chi_{r, 5}\bar{\chi}_{r, 9}\notag\\
    &+ \omega\chi_{r, 6}\bar{\chi}_{r, 2} + (-1 - e^{\frac{5}{6}\pi i})\chi_{r, 6}\bar{\chi}_{r, 4} + (-1 + \sqrt{3})\chi_{r, 6}\bar{\chi}_{r, 6} + (-1 - e^{\frac{5}{6}\pi i})\chi_{r, 6}\bar{\chi}_{r, 8} + \omega\chi_{r, 6}\bar{\chi}_{r, 10}\notag\\
    &+ (-1 + \omega)\chi_{r, 7}\bar{\chi}_{r, 3} + (-2 + 2\sqrt{3})\chi_{r, 7}\bar{\chi}_{r, 7} + \omega^2\chi_{r, 7}\bar{\chi}_{r, 9}\notag\\
    &+ (-1 + \sqrt{3})\chi_{r, 8}\bar{\chi}_{r, 4} + (-1 - e^{\frac{7}{6}\pi i})\chi_{r, 8}\bar{\chi}_{r, 6} + (-1 - i)\chi_{r, 8}\bar{\chi}_{r, 10}\notag\\
    &+ (-1 + \omega^2)\chi_{r, 9}\bar{\chi}_{r, 5} + \omega\chi_{r, 9}\bar{\chi}_{r, 7} + (1 + \sqrt{3})\chi_{r, 9}\bar{\chi}_{r, 9}\notag\\
    &+ (1 + \sqrt{3})\chi_{r, 10}\bar{\chi}_{r, 2} + \omega^2\chi_{r, 10}\bar{\chi}_{r, 6} + (- 1 + i)\chi_{r, 10}\bar{\chi}_{r, 8}\}.\label{142zu4}
\end{align}

At this point, it is instructive to check the consistency of the computation (\ref{142zu4}) with the F-move constraint  (\ref{constraintsf}). The work \cite{Thorngren:2021yso} showed that $(\hat{L}_{(1, 2, 2)})_{L_{(1, 2, 2)}}^{(r, s)(r^{\prime}, s^{\prime})}$'s in (\ref{122zu4}) are the solutions of (\ref{constraintsf}) for $L_x = L_y = L_z = L_{(1, 2, 2)}$. From this, we can do the same computation by using the fact that $L_{(1, 2, 2)}$ and $L_{(1, 4, 2)}$ are related via the involution. Since the F-move constraint can be exchanged under the involution, we conclude that $(\hat{L}_{(1, 4, 2)})_{L_{(1, 4, 2)}}^{(r, s)(r^{\prime}, s^{\prime})} = (\hat{L}_{(1, 2, 2)})_{L_{(1, 2, 2)}}^{(r, s)(r^{\prime}, \zeta(s^{\prime}))}$ are the solutions of (\ref{constraintsf}) for $L_x = L_y = L_z = L_{(1, 4, 2)}$.


Summing the above diagrams with the appropriate weights (\ref{m}), we finally obtain
\begin{align}
    Z_{\mathcal{M}(A_{10}, E_6)/\mathcal{A}_2} =& \dfrac{1}{3 + \sqrt{3}}\ \begin{tikzpicture}[scale=0.6,baseline={([yshift=-.5ex]current bounding box.center)},vertex/.style={anchor=base,
    circle,fill=black!25,minimum size=18pt,inner sep=2pt},scale=0.50]        
    \draw(0,0)--(6, 0)--(6,6)--(0,6)--(0,0);
    \draw[dotted,->](3,0)--(3.5,1);
    \draw[dotted](3.5,1)--(4,2);
    \draw[dotted,->](6,3)--(5,2.5);
    \draw[dotted](5,2.5)--(4,2);
    \draw[dotted,->](4,2)--(3,3);
    \draw[dotted](3,3)--(2,4);
    \draw[dotted,->](2,4)--(1,3.5);
    \draw[dotted](1,3.5)--(0,3);
    \draw[dotted,->](2,4)--(2.5,5);
    \draw[dotted](2.5,5)--(3,6);
    \end{tikzpicture}
    + \dfrac{1}{3 + \sqrt{3}}\ \begin{tikzpicture}[scale=0.6,baseline={([yshift=-.5ex]current bounding box.center)},vertex/.style={anchor=base,
    circle,fill=black!25,minimum size=18pt,inner sep=2pt},scale=0.50]        
    \draw(0,0)--(6, 0)--(6,6)--(0,6)--(0,0);
    \draw[blue, ->](3,0)--(3.5,1);
    \draw[blue](3.5,1)--(4,2);
    \draw[dotted,->](6,3)--(5,2.5);
    \draw[dotted](5,2.5)--(4,2);
    \draw[blue, ->](4,2)--(3,3);
    \draw[blue](3,3)--(2,4);
    \draw[dotted,->](2,4)--(1,3.5);
    \draw[dotted](1,3.5)--(0,3);
    \draw[blue, ->](2,4)--(2.5,5);
    \draw[blue](2.5,5)--(3,6);
    \end{tikzpicture}
    + \dfrac{1}{3 + \sqrt{3}}\ \begin{tikzpicture}[scale=0.6,baseline={([yshift=-.5ex]current bounding box.center)},vertex/.style={anchor=base,
    circle,fill=black!25,minimum size=18pt,inner sep=2pt},scale=0.50]        
    \draw(0,0)--(6, 0)--(6,6)--(0,6)--(0,0);
    \draw[dotted,->](3,0)--(3.5,1);
    \draw[dotted](3.5,1)--(4,2);
    \draw[blue, ->](6,3)--(5,2.5);
    \draw[blue](5,2.5)--(4,2);
    \draw[blue, ->](4,2)--(3,3);
    \draw[blue](3,3)--(2,4);
    \draw[blue, ->](2,4)--(1,3.5);
    \draw[blue](1,3.5)--(0,3);
    \draw[dotted,->](2,4)--(2.5,5);
    \draw[dotted](2.5,5)--(3,6);
    \end{tikzpicture}\notag\\
    &+ \dfrac{1}{3 + \sqrt{3}}\ \begin{tikzpicture}[scale=0.6,baseline={([yshift=-.5ex]current bounding box.center)},vertex/.style={anchor=base,
    circle,fill=black!25,minimum size=18pt,inner sep=2pt},scale=0.50]        
    \draw(0,0)--(6, 0)--(6,6)--(0,6)--(0,0);
    \draw[blue, ->](3,0)--(3.5,1);
    \draw[blue](3.5,1)--(4,2);
    \draw[blue, ->](6,3)--(5,2.5);
    \draw[blue](5,2.5)--(4,2);
    \draw[dotted,->](4,2)--(3,3);
    \draw[dotted](3,3)--(2,4);
    \draw[blue, ->](2,4)--(1,3.5);
    \draw[blue](1,3.5)--(0,3);
    \draw[blue, ->](2,4)--(2.5,5);
    \draw[blue](2.5,5)--(3,6);
    \end{tikzpicture}
    + \dfrac{\sqrt{3} - 1}{3 + \sqrt{3}}\ \begin{tikzpicture}[scale=0.6,baseline={([yshift=-.5ex]current bounding box.center)},vertex/.style={anchor=base,
    circle,fill=black!25,minimum size=18pt,inner sep=2pt},scale=0.50]        
    \draw(0,0)--(6, 0)--(6,6)--(0,6)--(0,0);
    \draw[blue, ->](3,0)--(3.5,1);
    \draw[blue](3.5,1)--(4,2);
    \draw[blue, ->](6,3)--(5,2.5);
    \draw[blue](5,2.5)--(4,2);
    \draw[blue, ->](4,2)--(3,3);
    \draw[blue](3,3)--(2,4);
    \draw[blue, ->](2,4)--(1,3.5);
    \draw[blue](1,3.5)--(0,3);
    \draw[blue, ->](2,4)--(2.5,5);
    \draw[blue](2.5,5)--(3,6);
    \end{tikzpicture}\notag\\
    = &\sum_{\substack{r = 1\\ \mathrm{step}\ 2}}^{9}\left(\sum_{s = 1, 3, 5, 6, 7, 9} \chi_{r, s}\bar{\chi}_{r, s} +  \chi_{r, 2}\bar{\chi}_{r, 10} + \chi_{r, 4}\bar{\chi}_{r, 8} + \chi_{r, 8}\bar{\chi}_{r, 4} + \chi_{r, 10}\bar{\chi}_{r, 2}\right)\notag\\
    = &Z_{\mathcal{M}(A_{10}, D_7)}.
\end{align}
Thus we conclude that the gauging of the algebra $\mathcal{A}_2$ results in the D-series minimal model.

Finally, we complete the ADE triality by establishing 
the reverse operation, namely, the gauging from $\mathcal{M}(A_{10}, D_7)$ to $\mathcal{M}(A_{10}, E_6)$. Let us first find the algebra. The quantum dimension of the algebra object which we have used for gauging from $\mathcal{M}(A_{10}, E_6)$ to $\mathcal{M}(A_{10}, D_7)$ is $\braket{\mathcal{A}_2} = 3 + \sqrt{3}$. From (\ref{AAdualqdim}), the quantum dimension of the algebra object $\mathcal{A}_2^{*}$ for gauging from $\mathcal{M}(A_{10}, D_7)$ to $\mathcal{M}(A_{10}, E_6)$ is also $3 + \sqrt{3}$. Recalling subsection \ref{D7TDL}, we find the algebra object satisfying this constraint must be\footnote{$1 + L_{(1, 5)}$ also satisfies the constraint, but we can exclude the possibility because it is not compatible with the involution.}
\begin{align}
    \mathcal{A}_2^{*} = 1 + L_{(1, 7)}. \label{reverse}
\end{align}
In comparison, let us also recall that the algebra object for gauging $\mathcal{M}(A_{10},A_{11})$ to obtain $\mathcal{M}({A_{10},E_{6}})$ is $\mathcal{A}_{E_6} = 1 + L_{(1, 7)}$ (which has the same label, but in different category).

The next step is to find the multiplication morphism and compute the diagrams. Since \eqref{reverse} is the binary algebra and its quantum dimension is the same as that of $\mathcal{A}_{E_6}$, the multiplication morphism is also the same as that used in $\mathcal{A}
_{E_6}$, whose explicit form can be found in \cite{Diatlyk:2023fwf}. The rest of the computation of gauging $\mathcal{A}_2^*$ is in parallel with gauging  $\mathcal{A}_{E_6}$ on  $\mathcal{M}(A_{10}, A_{11})$ in the similar way that gauging $\mathcal{A}_2$ and $\mathcal{A}_{E_6}^*$ was in parallel. The effect of $L_{(1, 7)}$ on $\mathcal{M}(A_{10}, D_{7})$ is different from $L_{(1, 7)}$ on $\mathcal{M}(A_{10}, A_{11})$, but it is related by the involution (\ref{eventwist}). See subsection \ref{D7TDL}. Thus, each diagram in the second line of (\ref{T/A1}) differs only in the even $s$ sector. However, the resulting partition function of $\mathcal{M}(A_{10}, E_6)$ after summing up five diagrams with the weight defined by the multiplication morphism is invariant under the involution, so the result of the non-invertible symmetry gauging by $\mathcal{A}_2^{*}$ is also $\mathcal{M}(A_{10}, E_{6})$.

\section{Discussions}\label{Conclusion}
In this paper, we have verified that D-series minimal models and $\mathrm{E}_6$-series minimal models are exchanged by non-invertible symmetry gauging. Combining the previous studies on gauging (non-)invertible symmetries, we have completed the picture of the ADE, more precisely AD$\mathrm{E}_6$, triality.

One application of the AD$\mathrm{E}_6$ triality from the non-invertible gauging is that the renormalization group flow discussed in \cite{Nakayama:2024msv} can be exported to the case of D-series and E-series (for integer $k$) because the line(s) used in the gauging commute with the relevant deformation used there.\footnote{This, for instance, makes the connection between \cite{Nakayama:2024msv} and \cite{Klebanov:2022syt,Katsevich:2024jgq} on a firmer footing because the latter primarily focus on the D-series flow.} On the other hand, the putative flow from the $(A_4, E_6)$ minimal model to the (tri)critical Ising model discussed in \cite{Nakayama:2022svf,Kikuchi:2024cjd} does not preserve the line used in the gauging, so the gauging will not help.  

One of the remaining tasks is to find the algebras that exchange A-series minimal models and $E_7(E_8)$-series minimal models. According to \cite{Bhardwaj:2017xup, Ostrik:2001xnt, Fuchs:2002cm, Kirillov:2001ti}, $\mathcal{A}_{E_7} = 1 + L_{(1, 9)} + L_{(1, 17)}$ implements the non-invertible symmetry gauging from $\mathcal{M}(A_{h - 1}, A_{17})$ to $\mathcal{M}(A_{h - 1}, E_7)$, and $\mathcal{A}_{E_8} = 1 + L_{(1, 11)} + L_{(1, 19)} + L_{(1, 29)}$ implements that from $\mathcal{M}(A_{h - 1}, A_{29})$ to $\mathcal{M}(A_{h - 1}, E_8)$. When we try to find their dual algebras and perform the explicit gauging, we may face several difficulties. The first is that we do not know whether the dual algebra objects are binary, but this may not be an obstruction because we have the constraint on their quantum dimensions. The second is that it is more non-trivial to determine their multiplication morphisms. In this paper, we have determined the multiplication morphisms easily by using the property that the algebra objects are binary. In contrast, for the algebra that is not necessarily binary, it may become more difficult to determine the multiplication morphisms. The third is that the more lines the dual algebra objects contain, the more diagrams we must compute in the actual gauging.  


Before concluding the paper, let us discuss several future directions to be pursued.

We have implemented the non-invertible symmetry gauging in continuous spacetime. It is also interesting to realize TDLs and gauge the non-invertible symmetries on the lattice \cite{Seifnashri:2025fgd}. Some TDLs in the crossed or direct channel are constructed in the RSOS models \cite{Chui:2002bp, Sinha:2023hum, Pearce:2024udz}, but to perform the gauging, we need to evaluate the partition function with the junctions of the TDLs with appropriate weight. The realization of such complicated configurations like (\ref{122zu3}) and (\ref{122zu4}) have not been studied in the literature.


In this paper, we have studied the torus partition function or spectrum of the theory obtained by gauging (non-)invertible symmetry. What about the operator product expansion (OPE) coefficients of the gauged theory? The recent conformal bootstrap analysis \cite{Nivesvivat:2025odb} suggests that once the spectrum is fixed, the OPEs must be given by the one proposed in \cite{Furlan:1989ra,Petkova:1994zs} uniquely fixed in all ADE cases, In principle, gauging on the higher genus will not only determine the spectrum but also OPE coefficients, and it should be interesting to pursue the idea.

It is also interesting to consider studying TDLs and gauging the non-invertible symmetry under the existence of the boundary. There is one interesting physical application here. Consider $(A_4, E_6)$ E-series minimal model, which is identified with the tensor product of the Ising model and Lee-Yang model. The boundary states in this theory correspond to the RG interface between the Ising model and the Lee-Yang model \cite{Konechny:2016eek}. The detailed study of the action of the TDLs on the different boundary conditions gives us information on how much symmetry is preserved under the renormalization group flow.

In this paper, we have restricted ourselves to gauging the (non-)invertible symmetries within the Virasoro minimal model. In principle, it is possible to combine the other topological field theories and gauge the combined symmetries. The ADE classification suggests that this would not lead to more modular invariant partition functions, but fermionic minimal was constructed in this way where the partition function depends on the spin structure. While the construction of \cite{Hsieh:2020uwb,Kulp:2020iet} is based on gauging the invertible symmetry, we should be able to obtain, for example, E-series fermionic minimal models from gauging non-invertible symmetries in the A-series bosonic minimal models or D-series fermionic minimal models from gauging non-invertible symmetries in the E-series bosonic minimal models.

Beyond Virasoro minimal models, gauging non-invertible symmetry has a huge potential that can generate novel CFTs unprecedented before. Once we know what is done can be undone, we can safely seek more opportunities in theoretical physics.

\section*{Acknowledgments}
The authors would like to thank C.~Luo, P.~Putrov, M.~You, and Y.~Wang for their helpful discussions.
TT is supported by JSPS KAKENHI Grant Number JP24KJ1500.
YN is in part supported by JSPS KAKENHI Grant Number 21K03581. 

\bibliography{ADEtriality}{}
\bibliographystyle{h-physrev}

\end{document}